\documentclass[rmp,preprint]{revtex4}
\usepackage{layout}

\usepackage[latin9]{inputenc}
\usepackage{amsmath}
\usepackage{amsfonts}
\usepackage{amssymb}
\usepackage{graphicx}
\usepackage{epsfig}

\begin{document}

\preprint{APS/123-QED}

\title{Destabilizing Taylor-Couette flow with suction}

\author{Basile Gallet}
\email{basile.gallet@ens.fr}
\affiliation{Laboratoire de physique statistique, Ecole Normale Sup\'erieure, 75231 Paris France}%
\author{Charles R. Doering}
\affiliation{Department of Mathematics, Department of Physics, and Center for the Study of Complex Systems, University of Michigan, Ann Arbor 48109-1043 USA}%
\author{Edward A. Spiegel}
\affiliation{Department of Astronomy, Columbia University, New York NY 10027 USA}

\date{\today}

\begin{abstract}
We consider the effect of 
radial fluid injection and suction on Taylor-Couette flow.  Injection at the 
outer cylinder and suction at the inner cylinder generally results in a 
linearly unstable steady spiralling flow, even for cylindrical shears that 
are linearly stable in the absence of a radial flux.  We study nonlinear
aspects of the unstable motions with the energy stability method.   
Our results, though specialized, may have implications for drag 
reduction by suction, accretion in astrophysical disks,  
and perhaps even in the flow in the earth's polar vortex.
\end{abstract}

\maketitle
\newcommand{\prt}[2]{\frac{\partial #1}{\partial #2}}
\newcommand{\sprt}[2]{\frac{\partial^2 #1}{\partial #2^2}}
\newcommand{\dsprt}[3]{\frac{\partial^2 #1}{\partial #2 \partial #3}}
\newcommand{\va}[1]{{\bf  #1}}
\newcommand{\intau}[1]{\int_{\tau}^{} #1 d\tau}

\section{The problem}

There are many contexts in which rotating flows arise and their instabilities are naturally of interest.  For an axisymmetric, incompressible, inviscid flow with no variation along the axis of symmetry, Rayleigh (1916) provided a necessary and sufficient condition for stability against axisymmetric perturbations (Chandrasekhar, 1961). 
Rayleigh also showed that a necessary criterion for the instability of such flows to non-axisymmetric perturbations is an analogue to his classic inflection point criterion for plane flows.  He expressed this criterion in words  but the correct mathematical expression is easily worked out following Rayleigh's lead (Spiegel and Zahn, 1970).   
For the study of the effect of viscosity on these results, the Taylor-Couette configuration of flow between coaxial cylinders provides a good proving ground (Chandrasekhar, 1961), although shear and rotation come together in other contexts as well (Yecko and Rossi, 2004 and references therein).

When criteria for the simplest two-dimensional circular shear flows indicate stability, the question naturally arises whether some simple extraneous effects may be destabilizing.   
In the case of the linearly stable plane shear flow, magnetic fields (Stern, 1963 and Chen and Morrison, 1991), Rossby waves (Lovelace {\it et al.}, 1999) and suction (Hocking, 1974; Doering {\it et al.}, 2000) have been found to be destabilizing mechanisms, given suitable conditions. 
In the case of  a stable Taylor-Couette flow, it had been seen even earlier that a magnetic field may be destabilizing (e.g. Chandrasekhar, 1961; Balbus and Hawley, 1998).  
In an attempt to add to this compendium of instabilities, we here investigate the instability that arises when a formerly circular flow is turned into a spiralling flow by the imposition of continous suction on the inner cylindrical boundary with a compensatory injection of fluid on the outer boundary.
For brevity, we shall hereinafter refer to the induced radial inflow as  {\it accretion} and the mechanism driving it as {\it suction}.    

A practical example of the conversion of differential circular flow to spiraling motion, or {\it swirl}, arises in the study of drag reduction on a cylindrical airfoil (Batchelor, 1967).
In that case,  when suction is imposed on the central cylindrical airfoil, the drag is reduced because boundary layer separation is inhibited.
Spiralling flows are also central to the structure and dynamics of accretion disks that form in a great variety of astrophysical situations such as binary stars (which are often X-ray sources), flows around massive black holes at the centers of galaxies, and in the disks around newly born stars where planetary systems are thought to form.   

Of course, accretion disks are more complicated than the simple model that we consider here.  In most astrophysical cases there is ionization so hydromagnetic effects  become important and can produce the magneto-rotational instability (see Balbus and Hawley, 1998).  However, protoplanetary nebulae such as the primitive solar nebula are believed to have been cool so that other mechanisms may need to be considered to get the full picture 
(for at least some phases) of the turbulent dynamics of planet formation.
The feature of the present study that may have relevance to the fluid dynamics of astrophysical disks is the radial inflow, or accretion that, in some astrophysical examples, is fed by inflow from an external mass source.
A similar combination of motions occurs also in the terrestrial polar vortex that may play a role in the transport of ozone in the upper atmosphere (McIntyre, 1995).

With these motivational examples in mind, we turn to an idealized pilot problem to learn what may happen when rotation and accretion combine to produce a spiralling motion as in Taylor-Couette flow with suction.
Although we cannot draw immediate conclusions about the applications of our results to the motivational examples just described, especially to accretion disks, we regard it as suggestive that the combination of rotation and accretion (i.e., radial injection and suction) in a flow leads to instability.
As we shall see in what follows, taking into account both accretion and rotation leads to very different stability properties of the flow than when only one of these effects alone is included.
For example, in the classical problem of the linear and nonlinear stability of Taylor-Couette flow between concentric cylinders,  the system is linearly stable when the outer cylinder is rotating and the inner one is stationary.
Likewise, simple radial accretion without rotation is linearly stable.
However, we find that, at sufficiently large rotational Reynolds numbers in a stable
TC flow, small accretion rates lead to linear instability.
Moreover we find that the rotational motion, rather than simple plane shear, plays a major role in the stability characteristics.  

An initially plane-parallel shear layer perturbed by transverse suction may also exhibit a small-suction instability (Hocking, 1974), while larger suction rates absolutely stabilize the corresponding laminar flow for arbitrarily large Reynolds numbers (Doering {\it et al.}, 2000).
This strong stabilizing effect of suction for even very high Reynolds numbers in the plane case does {\it not} carry over to the rotational case; in the cylindrical geometry transient growth of perturbations is generally encouraged by strong suction.

The following discussion of these issues is organized as follows. 
In section \ref{secform} we describe Taylor-Couette flow with suction. 
Then we treat the linear stability of this flow in section \ref{seclin} where we find linear instabilities in the simple flow of the model problem.
In section \ref{secnarrow} the asymptotic behavior of the onset of linear instability is computed for different limits in the narrow-gap approximation.
We next go on to consider some nonlinear aspects of the instability.
First, in section \ref{secenergy} we describe the energy stability method and then derive some rigorous bounds on the energy stability limit of the flow in section \ref{secbounds}.
That  limit is computed numerically in section \ref{secnumc}.
The concluding section \ref{conclusion} summarizes the results and mentions some possible implications.
In an appendix we note that physically relevant bounds on the mean energy dissipation rate for non-steady flows (including statistically stationary turbulence) remain unknown in the Taylor-Couette geometry with suction. 
 
\section{Formulation of the problem}
\label{secform}
\subsection{Geometry of the problem and boundary conditions}

Consider the incompressible flow of a Newtonian fluid between two coaxial porous cylinders.   For definiteness, we let the inner cylinder be at rest while the outer cylinder  is rotating with constant angular velocity, $\Omega$.  The radii of the inner and outer cylinders are, respectively, $R_1$ and $R_2$, and we use cylindrical 
coordinates $(\tilde{r},\theta,\tilde{z})$. 
We consider periodic boundary conditions in $\tilde{z}$ with period $L_z$, and impose no-slip boundary conditions so that the vertical ($\tilde{z}$) component of the velocity field vanishes at both cylinders and the azimuthal component matches the velocity of each of the cylinders.

\begin{figure}[h]
\begin{center}
\includegraphics[width=60 mm]{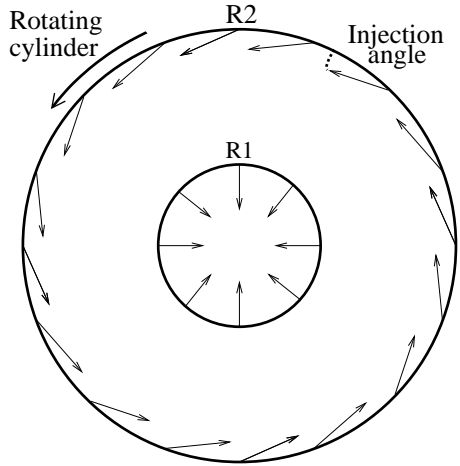}
\caption{The outer cylinder rotates at angular velocity $\Omega$.  Fluid is injected at the outer boundary with an entry angle $\Theta = \arctan[\varphi/(R_2^2\Omega)]$, and removed uniformly on the surface of the inner cylinder.}
\label{F1}
\end{center}
\end{figure}

Now we add a radial inward flow to this classical Taylor-Couette configuration.
Fluid is injected at the outer cylinder with the constant volume flux $2\pi\varphi$ (volume of fluid injected per unit time and per unit height of cylinder) and uniformly sucked out at the inner cylinder at the same rate.  The control parameter $\varphi \ge 0$ is central in what follows.  The resulting velocity field is ${\bf \tilde{u}}=\tilde{u}{\bf  e_r}+\tilde{v}{\bf  e_\theta}+\tilde{w}{\bf  e_z}$ and the boundary conditions are
\begin{eqnarray}
\label{BC1}
{\bf  \tilde{u}}(R_1,\theta,\tilde{z}) &=& -\frac{\varphi}{R_1} {\bf  e_r}\\
\label{BC2} {\bf  \tilde{u}}(R_2,\theta,\tilde{z}) &=& -\frac{\varphi}{R_2} {\bf  e_r} + R_2 \Omega {\bf  e_\theta}.
\end{eqnarray}
The incompressibility constraint $\va{\nabla}.\va{\tilde{u}}=0$ implies two further boundary conditions:
\begin{eqnarray}
 \tilde{u}_{\tilde{r}}(R_1)=0 \mbox{ and }  \tilde{u}_{\tilde{r}}(R_2)=0
\end{eqnarray}
where the subscript $\tilde{r}$ denotes partial differentiation with respect to  $\tilde{r}$. 
A schematic of the set-up is shown in Figure \ref{F1}.

\subsection{Dimensionless numbers}

This problem involves five parameters: the two radii, $R_1$ and $R_2$, the angular velocity of the outer cylinder, $\Omega$, the imposed accretion flux,  $2\pi \varphi$, and the viscosity of the fluid, $\nu$.
Nondimensionalizing time and space leaves three independent dimensionless numbers $\eta$,  $\Theta$ and $Re$ describing the geometry and physical boundary conditions:
\newline

\begin{itemize}
\item The geometrical factor is $\eta=\frac{R_1}{R_2}$. 
When $\eta \rightarrow 1$ we approach the narrow gap limit where $(R_2-R_1)<<R_1$, and expect to find results similar to those from the slab geometry --- namely plane Couette flow with suction.  On the other hand, when $\eta$ goes to zero, the outer radius $R_2$ goes to infinity relative to the inner radius and we expect to see significant effects of the curvature on the stability of the flow. 
\newline

\item The accretion flux is measured by the injection angle, $\Theta$, at the outer cylinder,
where
\begin{eqnarray}
\tan\Theta=\frac{\varphi}{R_2^2 \Omega}.
\end{eqnarray}
If $\tan \Theta = 0$, there is no suction and we recover a classical Taylor-Couette problem with a linearly stable stationary flow when the inner cylinder is stationary.
When $\tan \Theta \rightarrow \infty$ we approach the limit where suction dominates rotation.
\newline

\item The azimuthal Reynolds number is
\begin{eqnarray}
Re=\frac{R_2 \Omega (R_2-R_1)}{\nu} = \frac{R_2^2 \Omega (1 - \eta)}{\nu}.
\end{eqnarray}
We choose this definition of the Reynolds number to match
the definition of the Reynolds number for the plane Couette flow
when $\eta \rightarrow 1$.
\newline

\item Introduction of an additional parameter, the radial Reynolds number 
$\alpha=\varphi/\nu$, is useful for making some of the equations more compact. 
This parameter is not independent but is linked to $Re$, $\eta$ and $\tan(\Theta)$ by the relation
\begin{eqnarray}
\label{defalpha} \alpha=\frac{Re \tan \Theta}{1-\eta}.
\end{eqnarray}
\end{itemize}

We define the dimensionless velocity ${\bf u} = u{\bf  e_r}+v{\bf  e_\theta}+w{\bf  e_z} = {\bf \tilde{u}} / (R_2 \Omega)$ and the dimensionless coordinates $r=\tilde{r}/R_2$
and $z=\tilde{z}/R_2$ and the time $t = \Omega \tilde t$ .
The velocity field satisfies the incompressible Navier-Stokes equations, written in dimensionless form as 
\begin{eqnarray}
\label{NS}
{\bf u}_t +{\bf u} \cdot {\bf \nabla} {\bf u} = -{\bf \nabla} p + \frac{1-\eta}{Re} \Delta {\bf u}
\end{eqnarray}
with boundary conditions
\begin{eqnarray}
\label{BC1ssdim}
{\bf  u}(\eta,\theta,z,t) &=& -\frac{\tan \Theta}{\eta} {\bf  e_r}\\
\label{BC2ssdim} {\bf  u}(1,\theta,z,t) &=& -{\tan \Theta} ~{\bf  e_r} + ~{\bf  e_\theta}\\
u_r(\eta,\theta,z,t) &=& 0\\
u_r(1,\theta,z,t) &=& 0
\end{eqnarray}

\subsection{The basic laminar solution}
\label{lamsol} 
\begin{figure}[h]
\begin{center}
\includegraphics[width=100 mm]{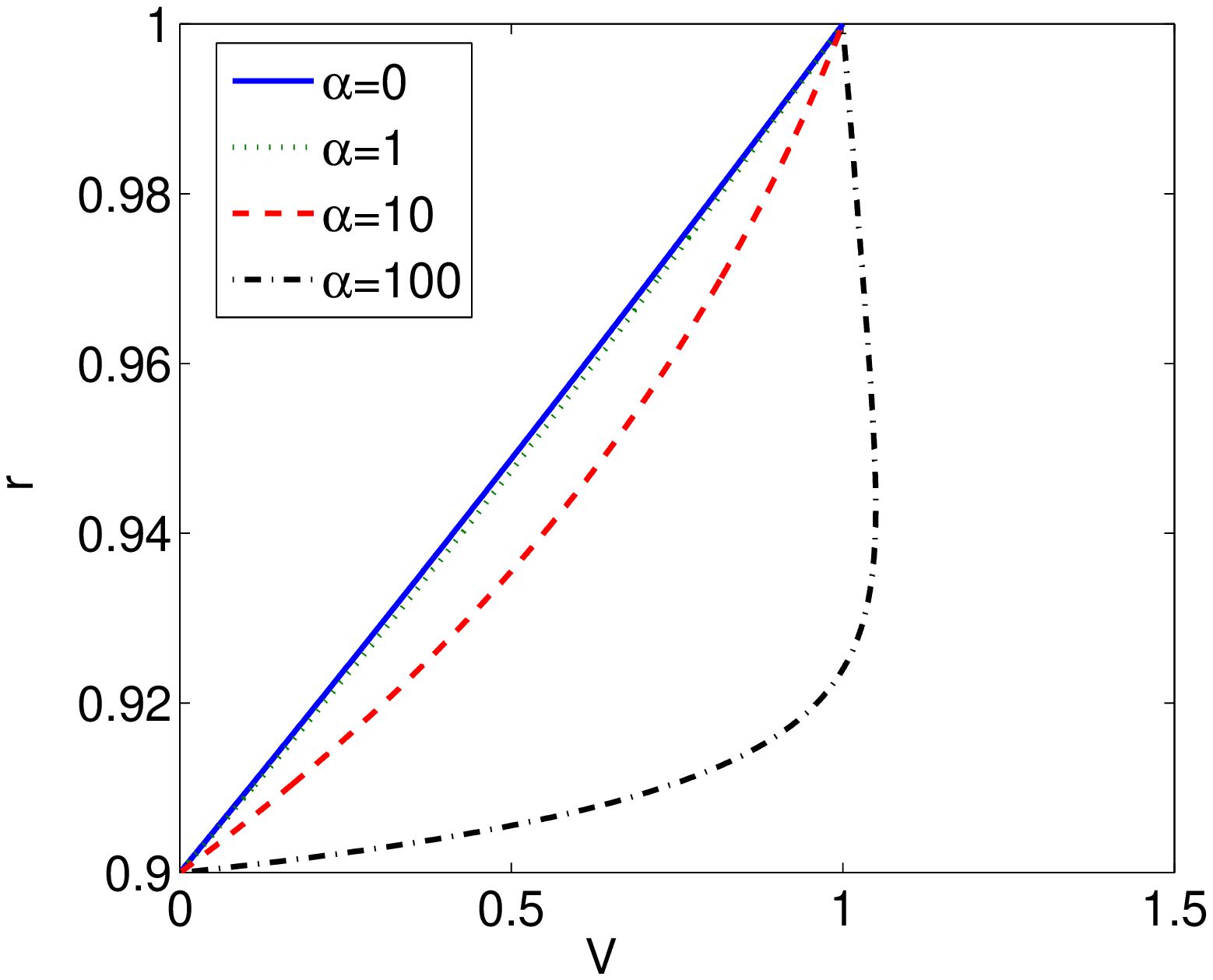}
\includegraphics[width=100 mm]{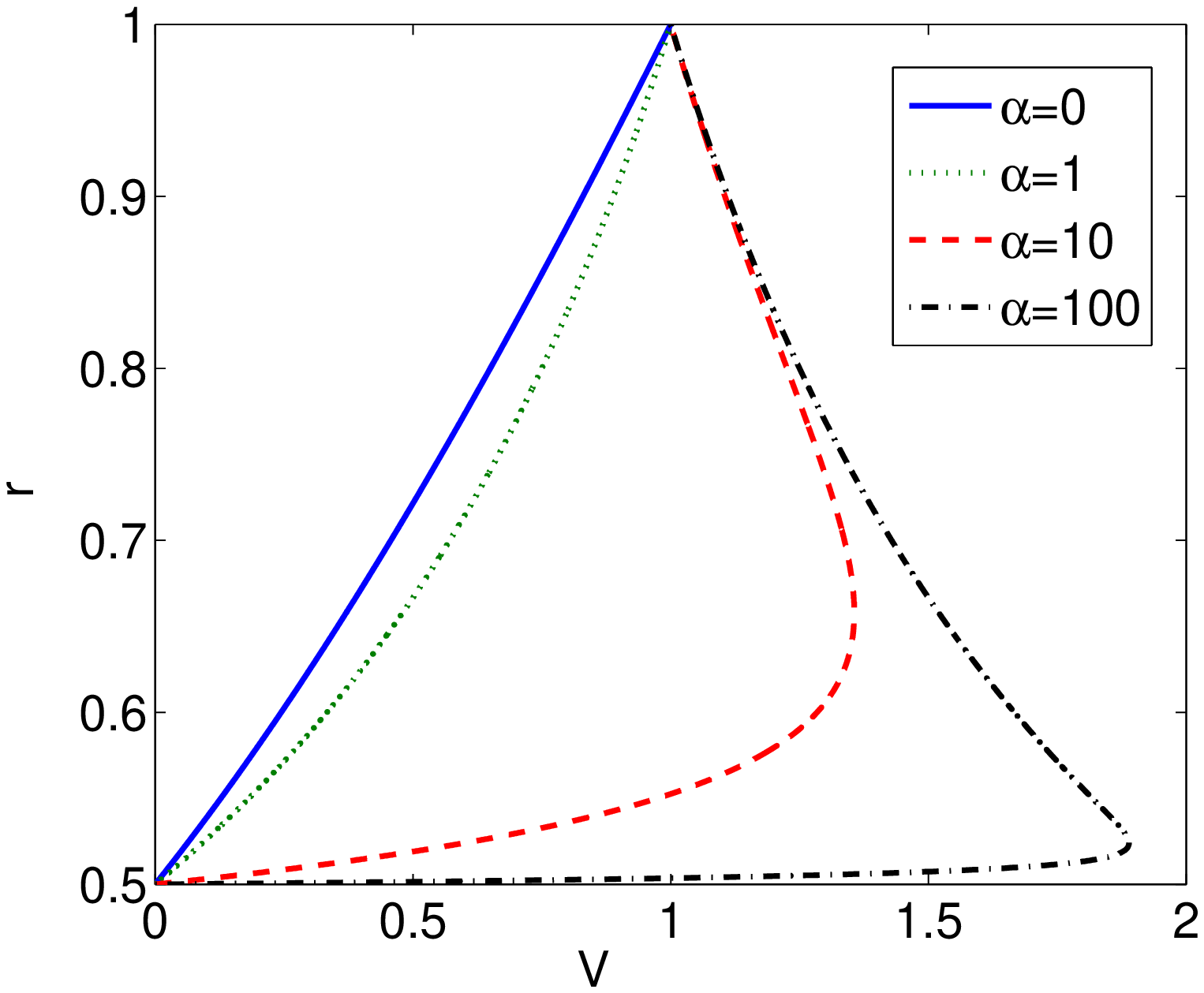}
\caption{Azimuthal velocity profiles for different values of the radial Reynolds number (top : $\eta=0.9$, bottom : $\eta=0.5$).}
\label{lamprof}
\end{center}
\end{figure}

The basic steady laminar solution ${\bf  V}_{lam}=(U(r),V(r),0)$ of equation (\ref{NS}) has radial velocity
\begin{eqnarray}
\label{radial}
U(r)=-\frac{\tan \Theta}{r}
\end{eqnarray}
and azimuthal velocity
\begin{eqnarray}
\label{azimuthal}
V(r)=\frac{1}{1-\eta^{2-\alpha}} r^{1-\alpha}+\frac{1}{1-\eta^{\alpha-2}} \frac{1}{r}\; .
\end{eqnarray}
The azimuthal velocity profile is represented in Figure \ref{lamprof} for several different combinations of values of $\alpha$ and $\eta$.

When $\alpha \rightarrow 0$ we recover the classical velocity field of the Taylor-Couette flow with a stationary inner cylinder.
That flow is independent of the kinematic viscosity, $\nu$, and the azimuthal velocity then increases monotonically when $r$ increases from $\eta$ to $1$.

For nonzero values of the entry angle $\tan \Theta$ and for sufficiently large values of the Reynolds number, that is, for $Re \tan \Theta >> 1$, the azimuthal speed increases outward (from the inner boundary) in a boundary layer of thickness $\delta$.
On the other hand, as the fluid comes in from the outer boundary, its motion is mostly inviscid and thus conserves angular momentum $r V(r)$: the azimuthal velocity of the fluid element increases during the inward journey from the outer boundary according to
\begin{eqnarray}
V(r) \approx \frac{1}{r}
\end{eqnarray}
For $\alpha>1$, the two flows join to form an azimuthal velocity profile with a maximum at $r=\eta+\delta$:
\begin{eqnarray}
V_r(\eta+\delta)=0 \ \ \Leftrightarrow \ \  \delta
=\eta[(\alpha-1)^{\frac{1}{\alpha-2}}-1].
\end{eqnarray}
This particular feature of the azimuthal velocity profile, that it can achieve a local maximum within the gap for $\alpha > 1$, is fundamentally linked to the rotational geometry.
This effect disappears in the plane limit $\eta \rightarrow 1$ since the maximum value of the azimuthal velocity inside the gap $V_{max}=\frac{1}{\eta}$.
For large values of $\alpha$, the boundary layer is very thin; ${\delta} \sim \eta \frac{ln(\alpha)}{\alpha}$.

These considerations point up a significant but not widely appreciated feature of accretion flows.
The asymptotic velocity profile as $Re \rightarrow \infty$ is dramatically modified from its structure at $\tan \Theta=0$ when $\tan \Theta \neq 0$.
When $\tan \Theta =0$, the azimuthal velocity profile is the  Reynolds-number-independent Taylor-Couette profile. 
When $\tan \Theta \neq 0$ the azimuthal velocity profile varies as $\frac{1}{r}$ everywhere inside the gap and outside the boundary layer, which becomes infinitely thin in the limit $Re \rightarrow \infty$. 
This shows that even if the accretion rate is very low (but not zero), one {\it cannot} neglect it at high $Re$ and study the problem with rotation only.
The limits $Re \rightarrow \infty$ and $\tan \Theta \rightarrow 0$ {\it do not commute}.

\section{Linear stability analysis and non-axisymmetric disturbances}
\label{seclin}
To study the destabilizing effect of the inward suction on the stability properties of Taylor-Couette flow, we perform a linear stability analysis.
It is already known from work by Min and Lueptow (1994) that, for axisymmetric disturbances to Taylor-Couette flow,  suction is stabilizing and so increases the critical Taylor number at which Taylor vortices first appear.
Therefore we here consider the linear stability theory for non-axisymmetric perturbations.
We continue to keep the inner cylinder stationary, so that the flow without suction is linearly stable according to Rayleigh's criteria for inviscid perturbations. 
 
In most instability mechanisms, only one symmetry of the initial problem is broken at the onset of the primary instability while others are broken through secondary instabilities.
Since we anticipate instability to nonaxisymmetric perturbations, we shall not, in the first instance, break the invariance to translations in the $z$-direction in seeking the primary instabilities of the spiralling flow.   We observe that, consistently with this supposition,
a Squire's Theorem exists in the plane-shear-with-suction situation, that is, in the limit $\eta \rightarrow 1$ (Doering {\it et al.}, 2000). 

\subsection{Linearization of the equations}

We make the decomposition $\va{u}=\va{V}_{lam}+\va{v}$ for the full velocity field, write the $r$, $\theta$, and $z$ components of a mode of the perturbation as
\begin{eqnarray}
\label{eqpert}
\va{v}=(u(r),v(r),0) e^{-\lambda t} e^{i m \theta}  
\end{eqnarray}
and seek the (possibly complex) eigenvalue $\lambda$.
The linearized versions of the Navier-Stokes equations for the $r$ and $\theta$ components of  $\va{v}$ for this mode are
\begin{eqnarray}
\lambda u &=&-\frac{1-\eta}{Re} u_{rr} - \frac{\tan \Theta + \frac{1-\eta}{Re}}{r} u_r + A_1 u + Z_1 v + p_r \label{lin1}\\
\lambda v &=& -\frac{1-\eta}{Re} v_{rr} - \frac{\tan \Theta +\frac{1-\eta}{Re}}{r} v_r + A_2 v + Z_2 u +
\frac{i m p}{r} \label{lin2}
\end{eqnarray}
where $p$ is the pressure perturbation and
\begin{eqnarray}
A_1 & = & i m \frac{V}{r} + \frac{\tan \Theta}{r^2} + \frac{1-\eta}{Re} \frac{m^2+1}{r^2}, \\
A_2 & = & i m \frac{V}{r} - \frac{\tan \Theta}{r^2} + \frac{1-\eta}{Re} \frac{m^2+1}{r^2}, \\
Z_1 & = & \frac{(1-\eta)}{Re} \frac{2 i m}{r^2} - \frac{2 V}{r}, \mbox{\ and}\\
Z_2 & = & V_r + \frac{V}{r} - \frac{(1-\eta)}{Re} \frac{2 i m}{r^2}.
\end{eqnarray}

The continuity equation becomes
\begin{equation}
\label{linmass}
i m v + (r u)_r = 0,
\end{equation}
providing an expression for $v$ that may be inserted into (\ref{lin2}) to give $p$ in terms of $u$.
Then, upon differentiating $p$ with respect to $r$ and inserting  the result into (\ref{lin1}), we arrive at a fourth order ODE for $u(r)$:
\begin{eqnarray}
\label{lin4thu}
& & \lambda \left[ -r^2 u_{rr} -3 r u_r + (m^2-1) u \right] = \left[\frac{(1-\eta)}{Re} r^2\right] u_{rrrr} \\
\nonumber & & + \left[\left(6 \frac{(1-\eta)}{Re} +\tan \Theta \right) r \right] u_{rrr} + \left[\frac{(1-\eta)}{Re} (6-m^2) + 3 
\tan \Theta  -r^2 A_2\right] u_{rr} \\
\nonumber & & + \left[-\frac{\tan \Theta+\frac{(1-\eta)}{Re}}{r}m^2 + i m r Z_1 +i m r Z_2 - r^2 (A_2)_r -3 r A_2 \right] u_r \\
\nonumber & & + \left[ m^2 A_1 + i m Z_1 + i m Z_2 + i m r (Z_2)_r - r (A_2)_r - A_2\right] u.
\end{eqnarray}
The boundary conditions are that $u$ and $u_r$ vanish at $r=\eta$ and $r=1$.

\begin{figure}[]
\begin{center}
\includegraphics[width=80 mm]{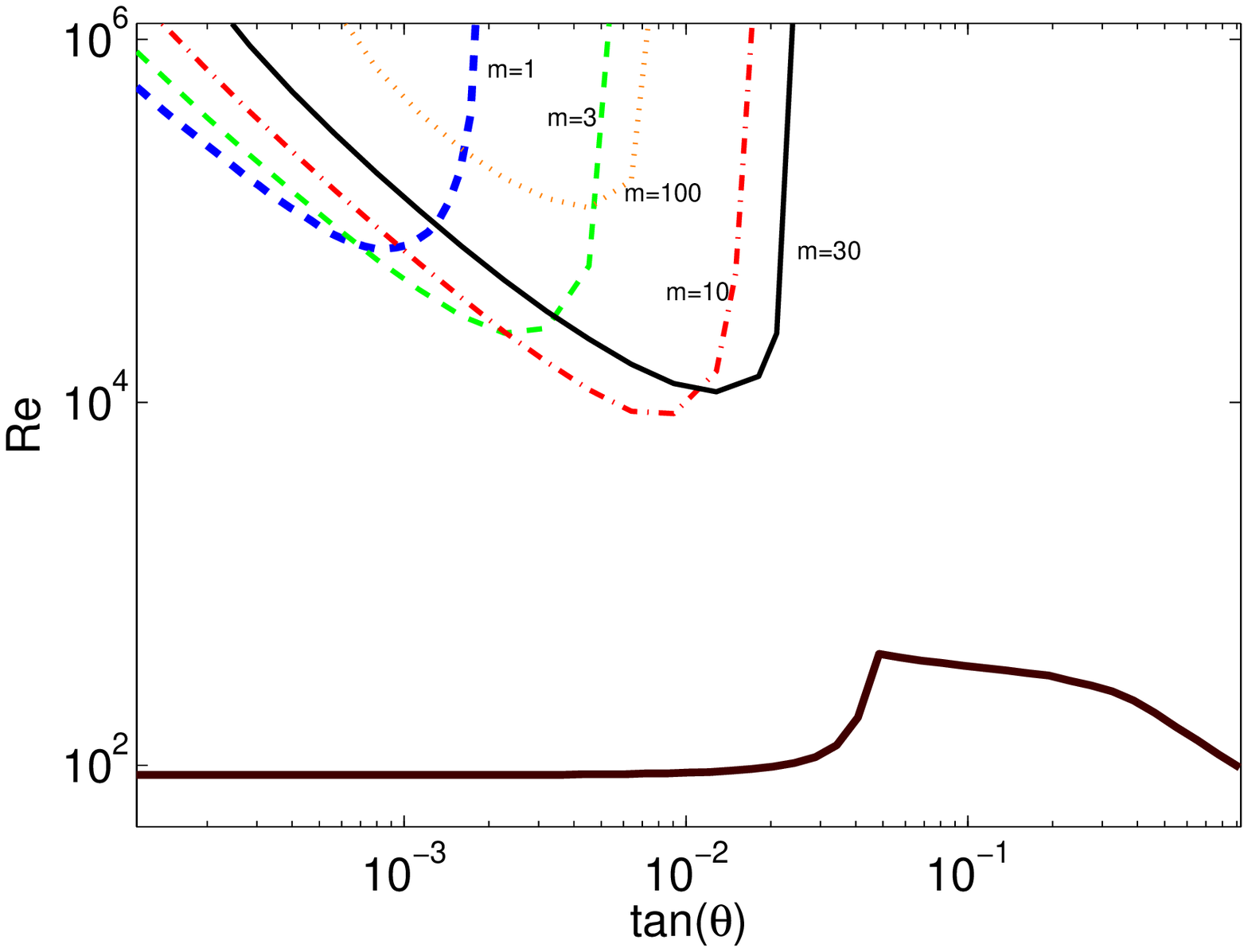}
\includegraphics[width=80 mm]{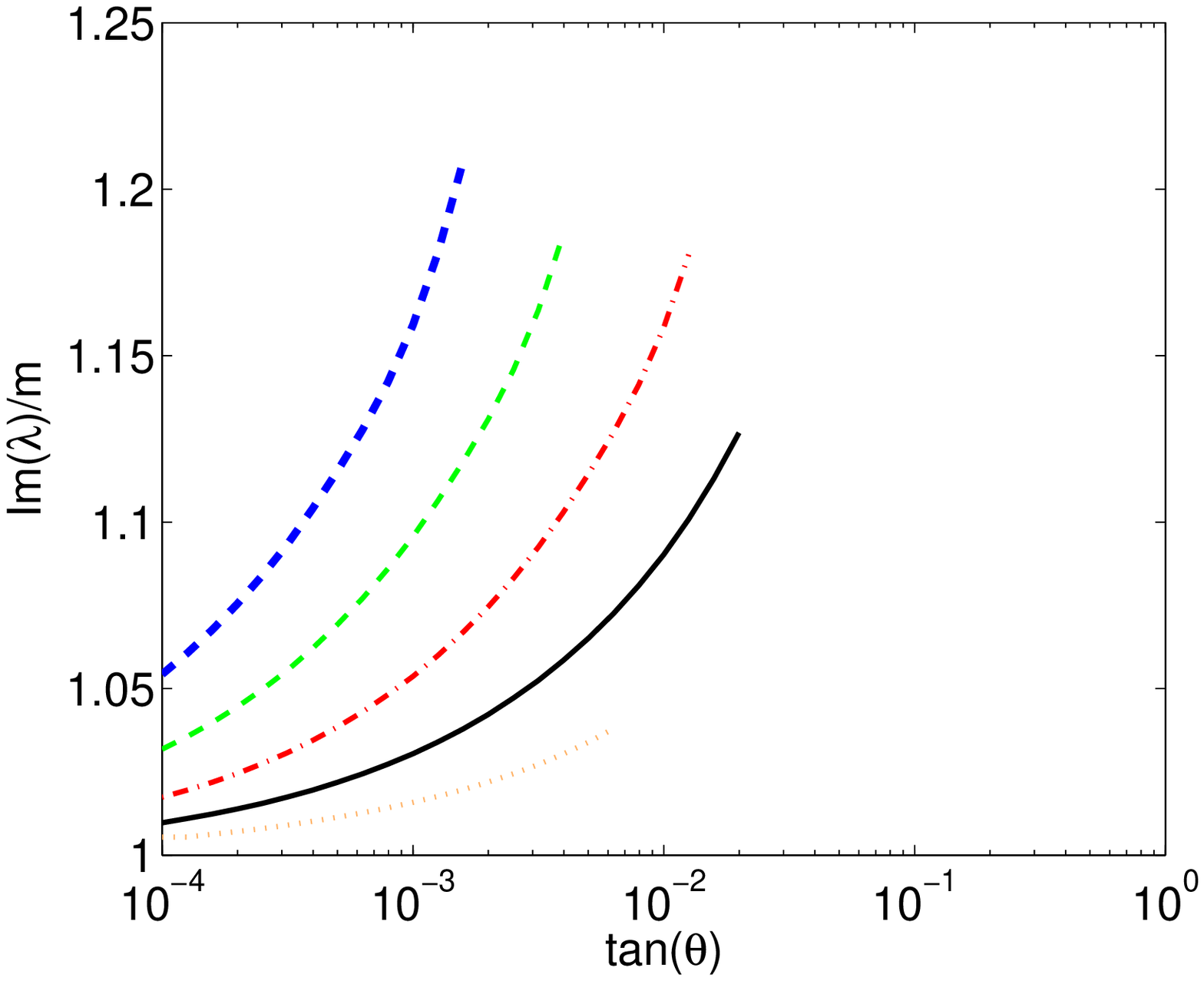}
\newline
\includegraphics[width=80 mm]{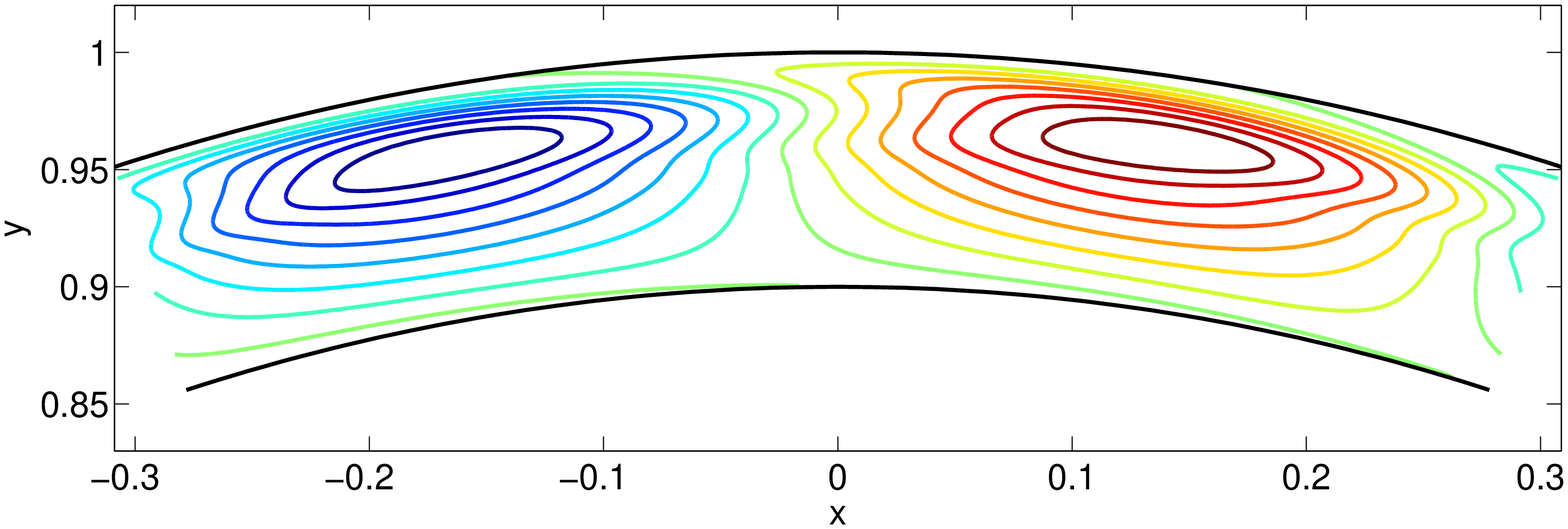}
\includegraphics[width=80 mm]{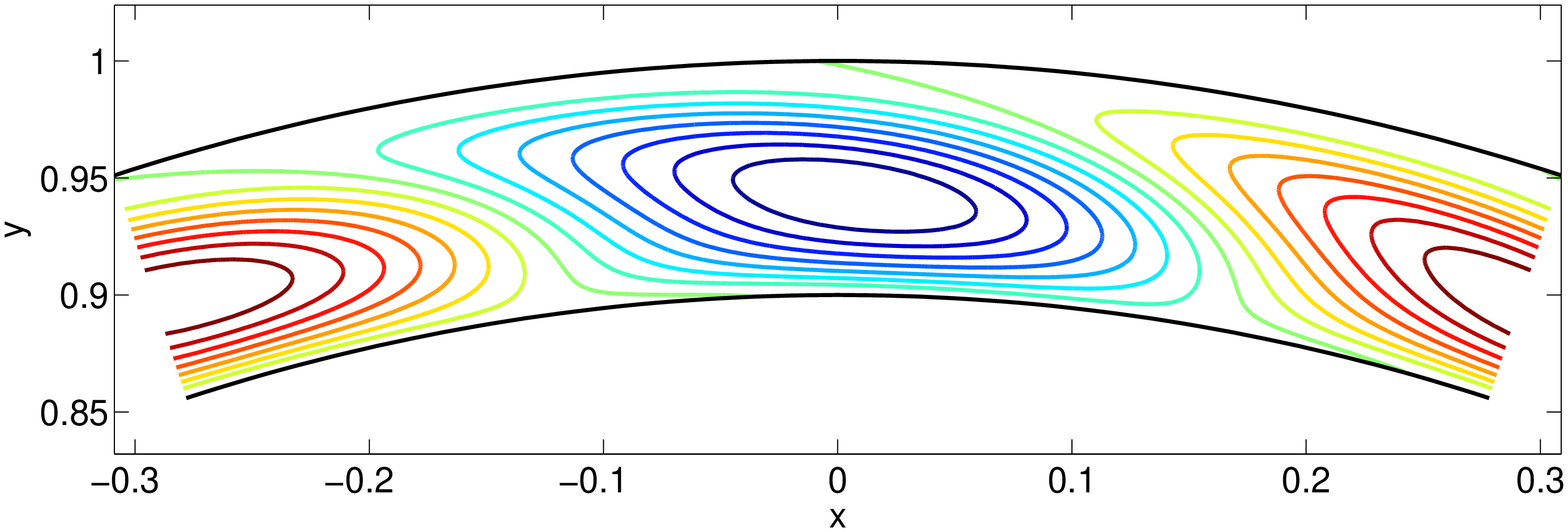}
\caption{\textbf{Top:} Linear stability limit of several modes for $\eta=0.9$ \textbf{(left)} and angular velocity of propagation of the unstable mode at onset \textbf{(right)}. Thick dashed line : $m=1$; dashed line : $m=3$; dash-dotted line : $m=10$; solid line : $m=30$; dotted line : $m=100$. The energy stability limit has been added to make the comparison easier (thick solid line).
\textbf{Bottom:} Contour plot of the streamfunction of the unstable mode at the onset of instability for $m=10$, $\eta=0.9$, and $\tan \Theta =0.001$ \textbf{(left)} and $\tan \Theta =0.01$ \textbf{(right)}. The smaller the injection angle, the more the velocity gradients are pressed up against the injection boundary.}
\label{lin0_9}
\end{center}
\end{figure}

\clearpage

\subsection{Numerical computation of the linear stability limit}

\begin{figure}[]
\begin{center}
\includegraphics[width=80 mm]{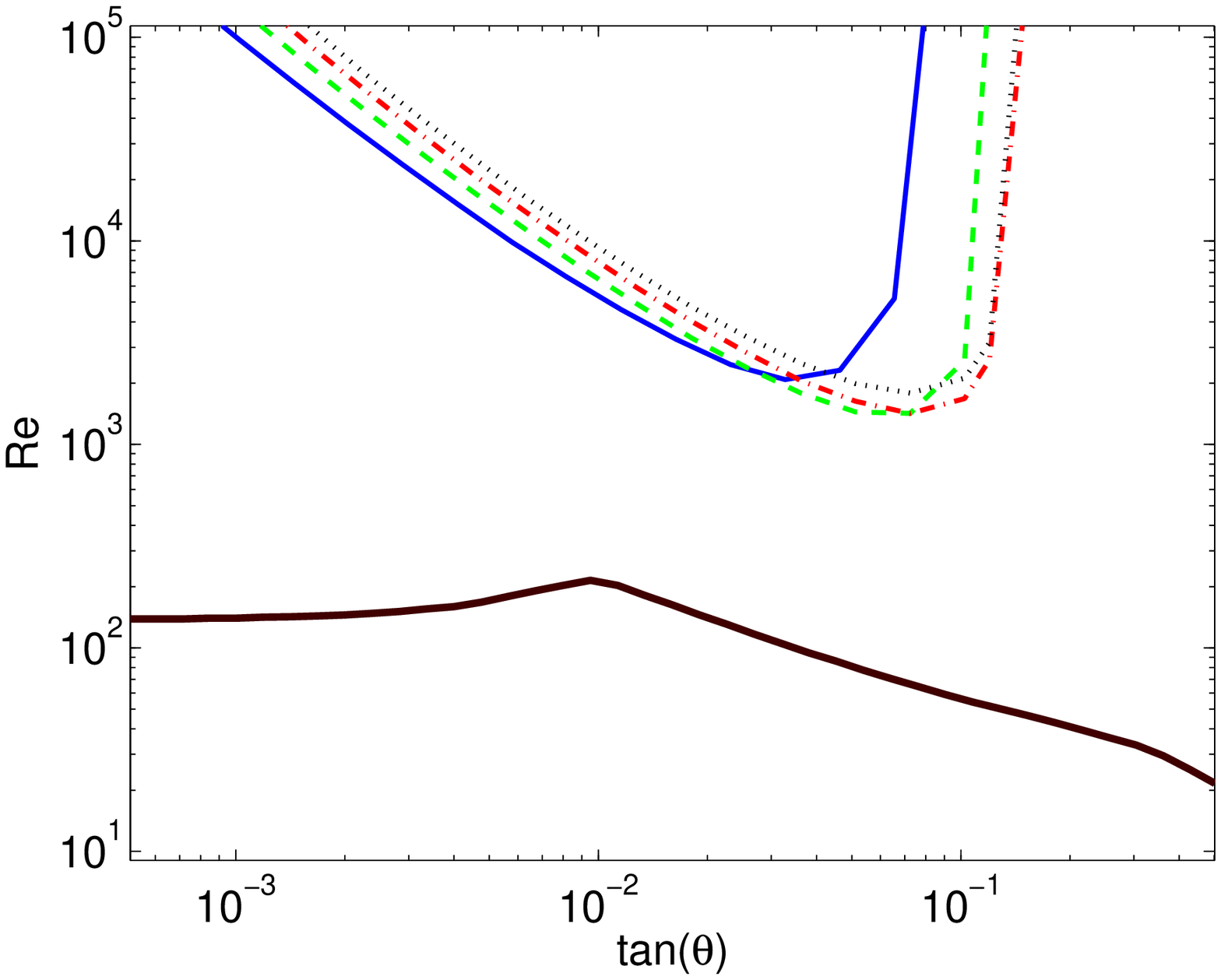}
\includegraphics[width=80 mm]{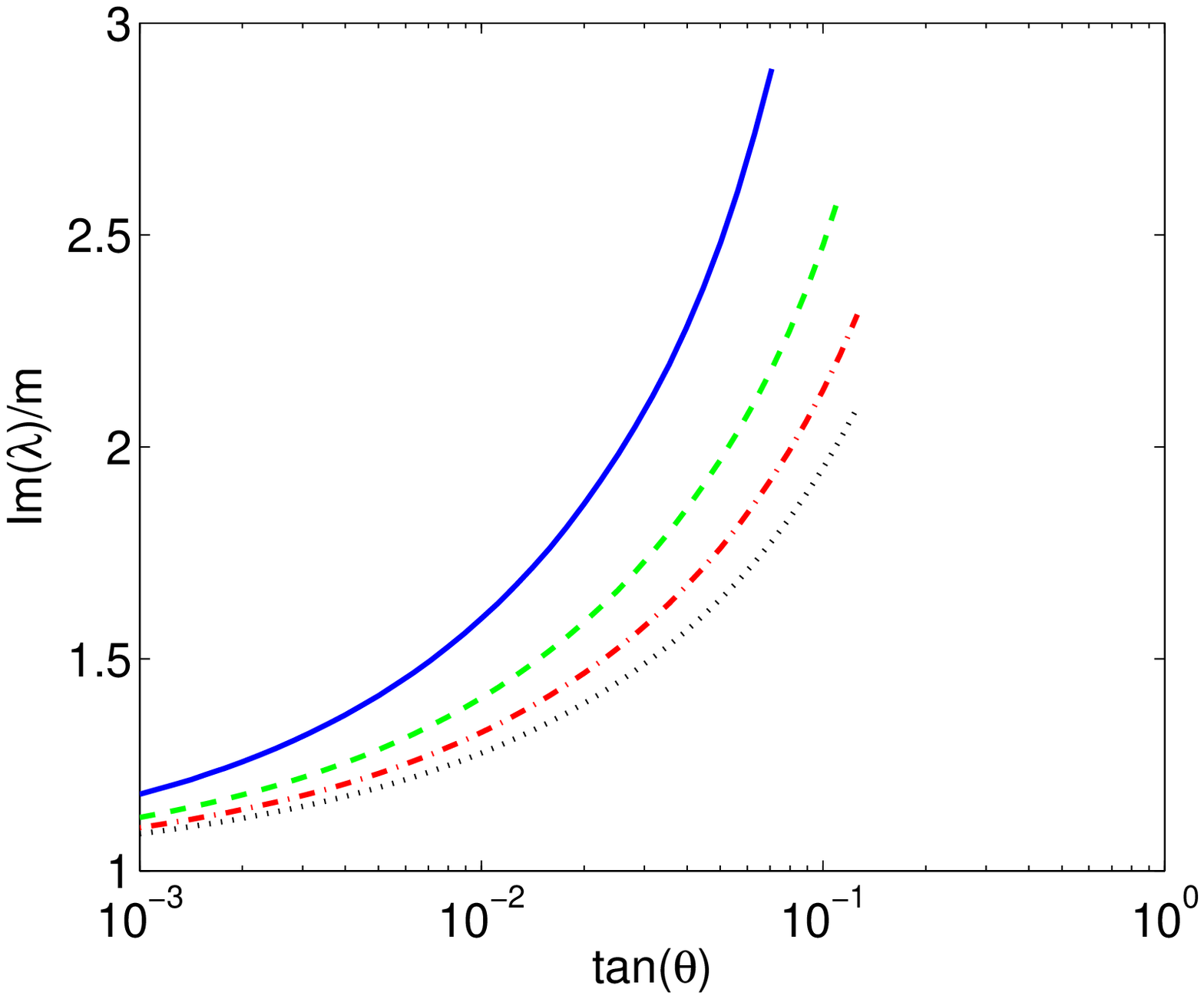}
\newline
\includegraphics[width=80 mm]{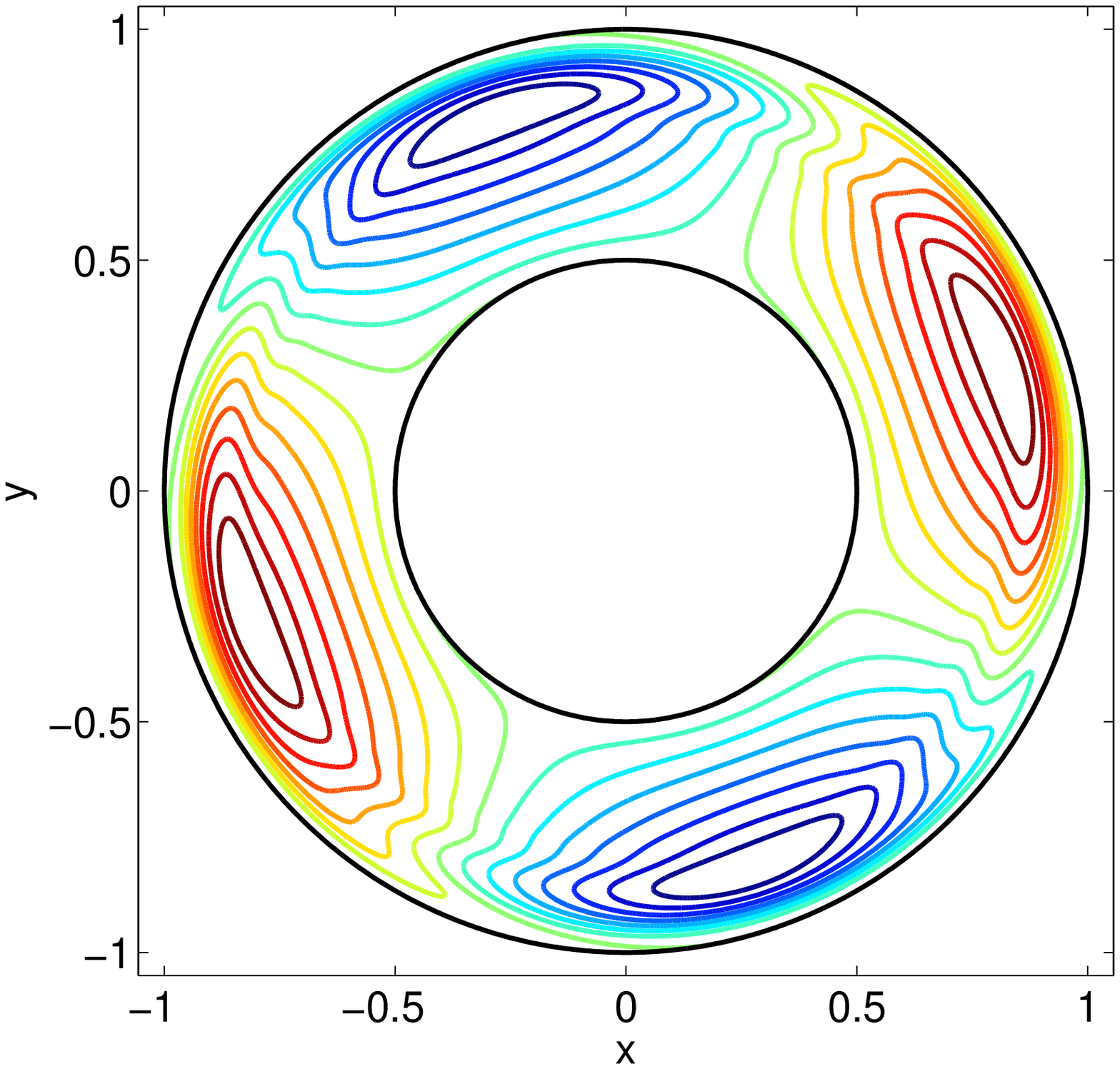}
\includegraphics[width=80 mm]{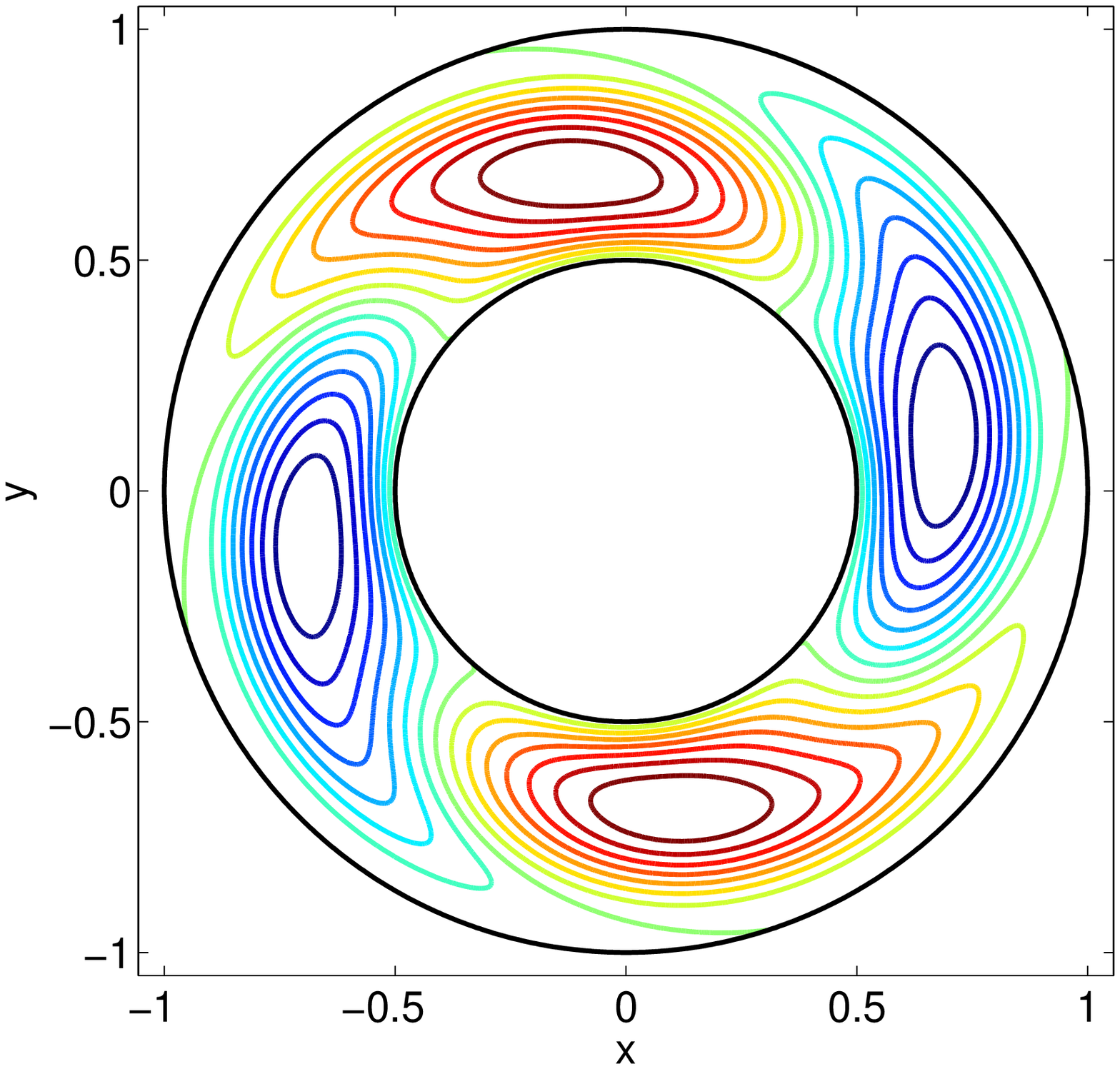}
\newline
\caption{\textbf{Top:} Linear stability limit of several modes for $\eta=0.5$ \textbf{(left)} and angular velocity of propagation of the unstable mode at onset \textbf{(right)}. In this high curvature geometry, linear instability is found for lower values of the Reynolds number and up to higher values of the injection angle. Solid line: $m=1$; dashed line: $m=2$; dashed-dotted line: $m=3$; dotted line: $m=4$; thick solid line: energy stability limit.
\textbf{Bottom left:} Contour plot of the streamfunction of the unstable mode at the onset of instability for $m=2$, $\eta=0.5$, and $\tan \Theta=0.005$.
The streamlines are concentrated toward the injection boundary.
\textbf{Bottom right:} Contour plot of the streamfunction of the unstable mode at the onset of instability for $m=2$, $\eta=0.5$ and $\tan \Theta=0.1$. The cells occupy the whole gap.}
\label{lin0_5}
\end{center}
\end{figure}

Equation (\ref{lin4thu}) presents an eigenvalue problem which can be solved numerically using a standard finite difference method.
The linear stability limit of different modes is represented in Figures \ref{lin0_9} and \ref{lin0_5} for $\eta=0.9$ and $\eta=0.5$.
There is linear instability even at small injection angles.
The modes that become unstable for the lowest critical Reynolds number have $m \sim \frac{1}{1-\eta}$, corresponding to closed and almost circular cells (in the $(r,\theta)$ plane) between the two cylinders. As expected for a system that lacks reflection symmetry in the $\theta$-direction, the unstable mode is a travelling wave that propagates in the azimuthal direction. The angular velocity of propagation $\mbox{Im}(\lambda)/m$ at the onset of instability is plotted in Figures \ref{lin0_9} and \ref{lin0_5}.

At a certain value of the injection angle, the linear stability boundary has a vertical asymptote in the $\tan \Theta-Re$ plane (see Figure \ref{lin0_9}).
Above this critical value of the injection angle, the flow is linearly stable at any Reynolds number.   Comparing the linear stability limits obtained for the two values of $\eta$, we notice that the flow is more unstable linearly for the geometry with the highest curvature.
From $\eta=0.9$ to $\eta=0.5$, the minimum value of the Reynolds number for marginal
stability --- the critical Reynolds number --- goes from approximately $10^4$ to $2 \times 10^3$ and the maximum value of the injection angle that allows a linear instability increases by almost an order of magnitude.
A few plots of the unstable modes at the onset of instability are presented in Figures \ref{lin0_9} and \ref{lin0_5}. 
Close to the critical injection angle the cells occupy the full width of the gap, whereas for smaller values of the injection angle the modes are pressed up against the injection boundary. In this latter situation, the angular velocity of propagation tends to 1, corresponding to the unstable mode being advected by the flow at the dimensionful angular velocity $\Omega$ of the outer boundary.

As already remarked, these linear instabilities exist when both rotation and accretion are present but they are lost if only one of these two ingredients is present in the configuration
considered here.

\section{Narrow-gap limit}
\label{secnarrow}
Since the numerical problem is readily solved in the linear case, we have the luxury of using some rough and ready approximation methods to  extract analytic results about linear stability on which one may build intuition about the solutions.
The numerical results provide guidance in the choice of effective approximations and the approximate solutions obtained serve to confirm those results.
We next carry out such approximations in the narrow-gap limit.
We consider situations where the critical Reynolds numbers are high enough that the laminar velocity profiles can be approximated throughout the gap by the their asymptotic forms for high Reynolds number,
\begin{equation}
\va{V_a}=\left\{
    \begin{array}{lll}
         - \tan \Theta/r\\
         1/r
          \ . \\
        \ \ \ \   0
    \end{array}
\right.
\end{equation}
This solution does not have the thin boundary layer of the large Reynolds number case.
Hence we do not satisfy the no-slip boundary condition on the azimuthal velocity at the inner boundary in this approximation. 
     
The vorticity, $\omega$, of a weak two-dimensional disturbance to this basic flow obeys the advection-diffusion equation,
\begin{equation}
\partial_t \va{\omega} + (\va{V_a}.\va{\nabla})\omega = \frac{1-\eta}{Re} \Delta \omega.
\end{equation}
The velocity perturbations (\ref{eqpert}) imply that the vorticity perturbation is of the form 
$\omega=\tilde{\omega}(r) e^{-\lambda t} e^{i m \theta}$.  Introduction of
this expression leads to
\begin{eqnarray}
\label{eqvorticity}
-\lambda \tilde{\omega} - \frac{1-\eta}{Re}\tilde{\omega}_{rr} - \left(\tan \Theta+\frac{1-\eta}{Re}\right) \frac{\tilde{\omega}_r}{r}+\left(\frac{i m}{r^2}+\frac{m^2}{r^2}\frac{1-\eta}{Re} \right)\tilde{\omega} = 0.
\end{eqnarray}
We introduce the variable $x=\frac{1-r/R_2}{1-\eta} \in [0,1]$ and define parameters
$S$, $\epsilon$ and $\mu$ according to
\begin{eqnarray}
& & S=m (1-\eta)^2 Re\\
& & \epsilon^2=\frac{\tan \Theta}{m (1-\eta)^2}\\
& & \lambda=im+{\mu}m(1-\eta).
\end{eqnarray}  
Equation (\ref{eqvorticity}) is then equivalent to:
\begin{eqnarray}
& & \frac{1}{S}\tilde{\omega}_{xx}-\left( \epsilon^2 +
\frac{1-\eta}{S}\right)\frac{\tilde{\omega}_{x}}{1-(1-\eta)x}\\
\nonumber & & +\left(-\frac{m^2 (1-\eta)^2}{S}+{\mu}[1-(1-\eta)x]^2 \right.
\\
\nonumber & & -ix(2-(1-\eta)x)\bigg{)}\frac{\tilde{\omega}}{(1-(1-\eta)x)^2}=0.
\end{eqnarray}  
In the narrow gap limit of this equation, for which $(1-\eta)x \ll 1$, we find 
\begin{eqnarray}
\label{eqomega}
\frac{1}{S}\tilde{\omega}_{xx}-\left( \epsilon^2 + \frac{1-\eta}{S}\right)\tilde{\omega}_x+\left(-\frac{m^2 (1-\eta)^2}{S}+{\mu}-2ix \right)\tilde{\omega}=0.
\end{eqnarray}  

\subsection{Linear stability limit for small injection angles}
\begin{figure}[]
\begin{center}
\includegraphics[width=80 mm]{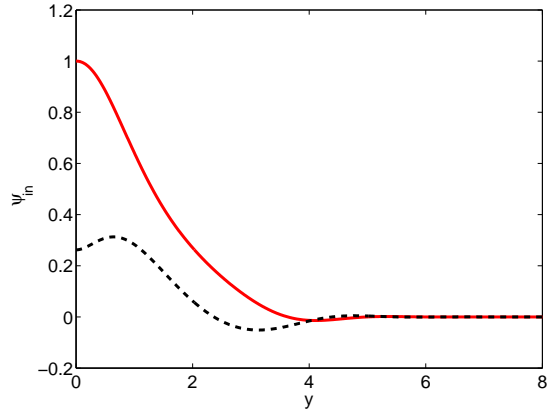}
\caption{Inner solution at the onset of instability $N=N_c=4.58$ and $a=5.62 i$ (solid line: real part, dashed line: imaginary part).}
\label{BLphi}
\end{center}
\end{figure}

As we saw, instability requires both rotation (the first term of (\ref{eqomega})), and accretion (the term proportional to $\epsilon^2$). 
Hence we require that the three terms of equation (\ref{eqomega})  are of the same order of magnitude within the active layer, the layer close to the injection boundary ($x=0$) where the main development of the unstable mode occurs  for small injection angles.
We bring this out by introducing  the scalings 
\begin{equation}
\label{scales}
S=\frac{N}{\epsilon^3}, \qquad {\mu}=\epsilon a \qquad {\rm and} \qquad
x=\epsilon y,
\end{equation} with $N$, $a$ and $y$ all ${\cal O}(1)$ inside the boundary layer. 
Equation (\ref{eqomega}) then becomes
\begin{eqnarray}
\label{eqscaling}
& & \frac{1}{N}\tilde{\omega}_{yy}-\tilde{\omega}_{y}+(a-2iy)\tilde{\omega}
={\cal O}(\epsilon)
\end{eqnarray}  
We are interested only in the leading order behavior in $\epsilon$, so we discard the right hand side, which is proportional to $\epsilon$. 
In the narrow-gap limit, the vorticity is linked to the streamfunction $\psi$ of the perturbation by
\begin{eqnarray}
\label{omegaphi}
\tilde{\omega} &\simeq& -\psi_{rr}+m^2\psi  = \frac{1}{(1-\eta)^2} \left( -\frac{\psi_{yy}}{\epsilon^2} + m^2 (1-\eta)^2 \psi \right).
\end{eqnarray}  
Hence, at the leading order in $\epsilon$, $\tilde{\omega}$ may
be replaced in (\ref{eqscaling}) by $\psi_{yy}$.
Thus the streamfunction close to the boundary satisfies
\begin{equation}
\frac{1}{N}\psi_{yyyy} - \psi_{yyy} +\left( a - 2iy \right)\psi_{yy} = 0.
\end{equation}
The solution for $\psi_{yy}$ that vanishes for large $y$ can be expressed in terms of the Airy function:
\begin{equation}
\psi_{yy}(y) = e^{\frac{N}{2}y} Ai \left(-\frac{N^{1/3} a}{(2 i)^{2/3}} + \frac{N^{4/3}}{4(2 i)^{2/3}}+(2 i)^{1/3}N^{1/3} y\right).
\end{equation}
If we let $\psi_{in}$ denote the corresponding solution that vanishes for large $y$, i.e., the inner solution, we can write the general solution as
\begin{equation}
\psi(y)=A y +B + \psi_{in}(y).
\end{equation} 
The boundary condition $\psi(\frac{1}{\epsilon})=0$ gives $A y +B=A (y-\frac{1}{\epsilon})$.
The boundary condition $\psi(0)=0$ leads to $A=\epsilon \psi_{in}(0)$  and, finally, the boundary condition $\psi_y(0)=0$ gives $\epsilon \psi_{in}(0) + (\frac{d \psi_{in}}{dy})_{y=0}=0$.
In the limit $\epsilon \rightarrow 0$ this relation becomes $(\frac{d \psi_{in}}{dy})_{y=0}=0$.
The parameter $a$ is then a solution of the implicit equation
\begin{eqnarray}
& &\int_{0}^{\infty} e^{\frac{N}{2}y} Ai \left(-\frac{N^{1/3} a}{(2 i)^{2/3}} + \frac{N^{4/3}}{4(2 i)^{2/3}}+(2 i)^{1/3}N^{1/3} y\right)dy = 0.
\end{eqnarray}
The onset of instability is found for $N=N_c$ so that $a$ is purely imaginary. 
We get the critical values $N_c \simeq 4.58$ and $a=5.62 i$. 
The asymptotic behavior for small values of the injection angle then corresponds to
\begin{eqnarray}
\label{asymptsmalli}
Re_c &\sim& 4.58 m^{1/2} (1-\eta) \tan \Theta^{-3/2}, \nonumber \\
\mbox{Im}(\lambda)&=& m + 5.62 \sqrt{m \tan(\theta)}.
\end{eqnarray}
The imaginary part of the eigenvalue corresponds to the advection of the pattern by the mean flow close to the outer boundary: for small injection angles, the eigenmode is a traveling wave of dimensionful angular velocity $\Omega$. 
The inner solution at the onset of instability is represented in Figure \ref{BLphi}.

\subsection{Critical injection angle for the smallest azimuthal wavenumbers}
\begin{figure}[]
\begin{center}
\includegraphics[width=80 mm]{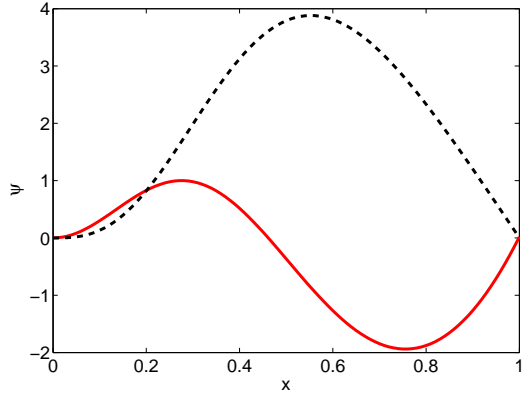}
\caption{Marginally stable mode at the onset of instability $\epsilon_c^2=0.166$ and ${\mu}=2.04 i$, for an infinite Reynolds number (solid line: real part, dashed line: imaginary part).}
\label{modetanthetac}
\end{center}
\end{figure}

Close to the critical injection angle for which the critical Reynolds number goes to infinity, the instability develops in the whole width of the gap. 
The parameter $\epsilon$ does not necessarily have to be small, but we simplify our calculations by focusing on the lowest of the azimuthal wavenumbers --- specifically on $m^2 (1-\eta)^2 \ll 1$. 
At the leading order in $m^2 (1-\eta)^2$, equation (\ref{omegaphi}) reduces to
\begin{equation}
\tilde{\omega}=-\frac{\psi_{xx}}{(1-\eta)^2}.
\end{equation}
Inserting this into (\ref{eqomega}) and taking the limit $S \rightarrow \infty$ gives
\begin{equation}
\epsilon^2 \psi_{xxx} + (2ix - {\mu})\psi_{xx}=0
\end{equation}
with solution
\begin{equation}
\psi_{xx}=\exp \left( \frac{1}{\epsilon^2} ({\mu}x-ix^2)  \right).
\end{equation}
Imposing the boundary conditions $\psi(0)=0$, $d_x(\psi)_{x=0}=0$, 
$\psi(1)=0$, 
we deduce the dispersion relation
\begin{equation}
\int_{t=0}^1\int_{x=0}^t \exp \left( \frac{1}{\epsilon^2} ({\mu}x-ix^2)  \right) dx dt = 0.
\end{equation}
Once again, the critical injection angle is obtained when the solution ${\mu}$ of this implicit equation has vanishing real part. 
This yields the values $\epsilon_c^2=0.166$ and ${\mu}=2.04 i$, and hence the asymptotic behavior is given by
\begin{eqnarray}
\label{vertasympt}
\tan \Theta_c &=& 0.166 m (1-\eta)^2 \nonumber \\
 \mbox{Im}(\lambda) &=& m + 2.04 m (1-\eta)
\end{eqnarray}
The unstable mode is shown in Figure \ref{modetanthetac}.

\subsection{Collapsing the different marginal stability curves}

\begin{figure}[]
\begin{center}
\includegraphics[width=90 mm]{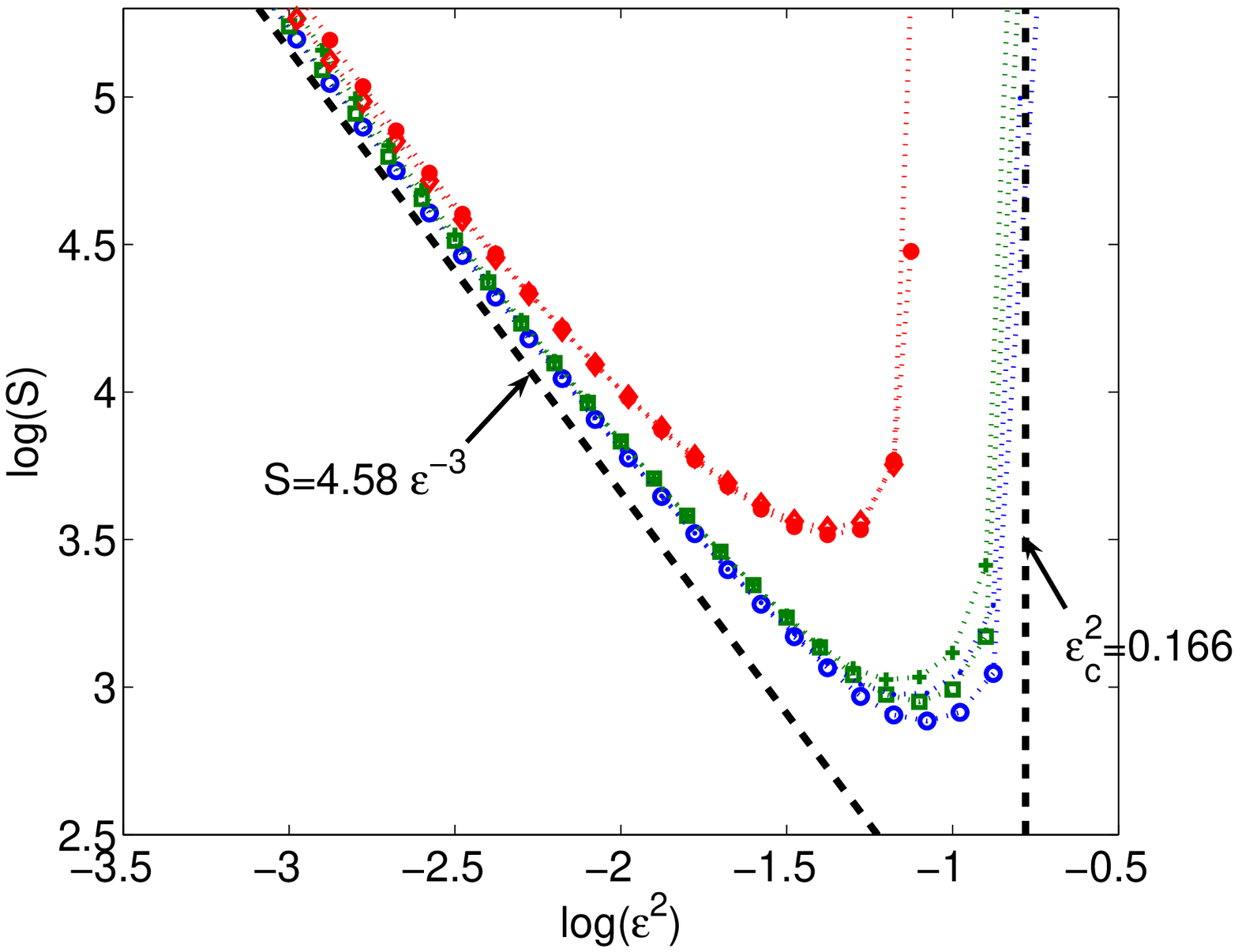}
\includegraphics[width=90 mm]{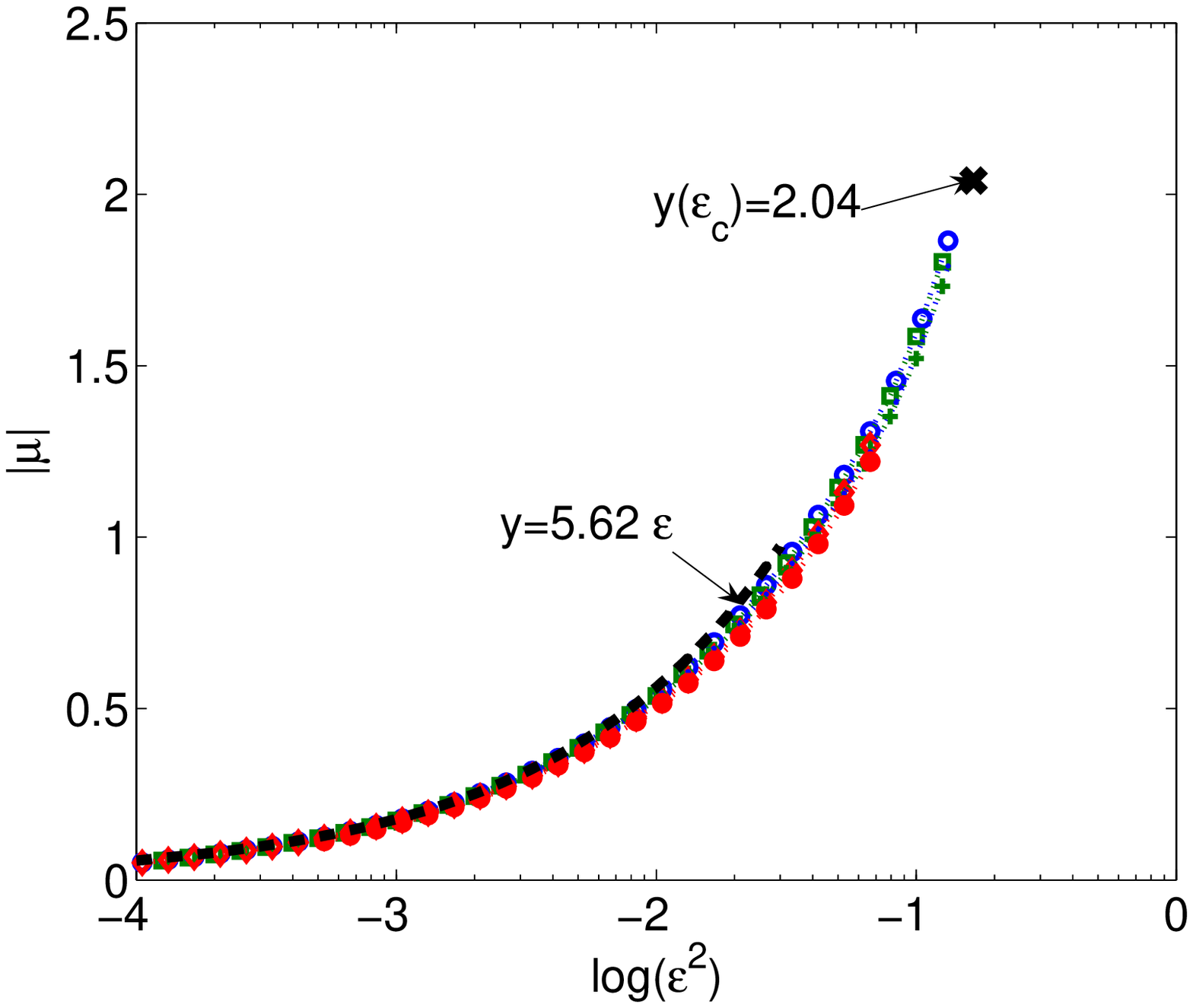}
\caption{\textbf{Top:} Curves of marginal stability for different values of $\eta$ and different azimuthal wavenumbers $m$, in the $\epsilon^2-\, S$ plane.
The small injection-angle asymptote is valid for every curve in the small gap limit.
There is always a vertical asymptote but its placement is accurate for small wavenumbers only ($m^2 (1-\eta)^2 << 1$).
\textbf{Bottom:} (Rescaled) deviation of the angular velocity of propagation to that of the outer cylinder. The asymptotic behaviors have been drawn for comparison.   
The symbols are $\circ$: $\eta=0.9$, $m=3$; $\square$: $\eta=0.9$, $m=10$; $\diamond$: $\eta=0.9$, $m=30$; $\cdot$: $\eta=0.99$, $m=30$; $+$: $\eta=0.99$, $m=100$; $\bullet$: $\eta=0.99$, $m=300$. }
\label{linstabR}
\end{center}
\end{figure}

The marginal stability curves are plotted on Figure \ref{linstabR} in the $\epsilon^2 -\, S$ plane for different values of $\eta$ (in the small gap limit) and different values of the azimuthal wavenumber $m$.
The curves corresponding to the lowest azimuthal wavenumbers (in the sense that $m^2 (1-\eta)^2 \ll 1$), collapse onto a single curve. 
For small injection angles, this curve admits an asymptote given by equation (\ref{asymptsmalli}). 
Its critical injection angle is given by the vertical asymptote derived in (\ref{vertasympt}). 
For high azimuthal wavenumbers, the asymptote to the limit of small injection angles is still acurate, but the vertical one is not.
This comes from the fact that one cannot simplify the relation $\omega \simeq -\frac{\psi_{xx}}{(1-\eta)^2} + m^2 \psi$, and thus cannot ignore the dimensionless parameter $m^2 (1-\eta)^2$ in the limit of high azimuthal wavenumbers.
The critical injection angle goes to zero in this limit, and its behavior may be determined with more careful asymptotics.

The angular velocity of propagation of the unstable modes are collapsed onto a single master curve in Figure \ref{linstabR} by plotting the quantity $|\mu|=\left(\frac{\mbox{Im}(\lambda)}{m}-1 \right)/(1-\eta)$ as a function of $\epsilon^2$. This quantity can be thought of as (a rescaled version of) the deviation of the angular velocity of propagation to the velocity of the outer boundary. It goes to zero as $\epsilon$ tends to zero following the scaling law (\ref{asymptsmalli}), whereas it tends to the finite value $2.04$ as $\epsilon \rightarrow \epsilon_c$ according to equation (\ref{vertasympt}).

\section{Energy stability of the basic flow}
\label{secenergy}
Energy stability is defined below; it is a strong form of nonlinear stability theory.
It provides sufficient conditions for a flow to be stable to perturbations of \textit{arbitrary} amplitude.
However a flow that is not energy stable may still be linearly stable.
In this section we formulate the problem of determining the energy stability limit of the base flow introduced above.
 
\subsection{Evolution equation for the energy of the deviation from the laminar solution}
\label{decomp} The starting point for the analysis is the decomposition of the velocity field ${\bf u}$ into the steady laminar flow and a time dependent deviation.
In this analysis the deviation from the laminar solution need not be small.
We thus introduce the decomposition ${\bf  u}({\bf  r},t)={\bf  V}_{lam}({{\bf  r}})+{\bf  v}({\bf  r},t)$ into equation (\ref{NS}) to obtain
\begin{eqnarray}
\label{NSdecomp}
\va{v}_t+(\va{v}.\va{\nabla})\va{v}&+&(\va{V}_{lam}.\va{\nabla})\va{v}+(\va{v}.\va{\nabla})\va{V}_{lam} = - \va{\nabla}p + \frac{1-\eta}{Re} \Delta \va{v} \; .
\end{eqnarray}
The time-dependent velocity field ${\bf  v}({\bf  r},t)$ is divergence free, satisfies the homogeneous boundary conditions ${\bf  v}={\bf  0}$ at the two cylinders, and is assumed to be periodic in $z$ with period $L_z$. 

To study the kinetic energy of this variable field we take the dot product of this equation with $\va{v}$ and integrate over one cell, i.e. over the domain $\tau=[R_1,R_2]\times[0,2\pi]\times[0,L_z]$. 
Let us denote the volume element in this domain by $d\tau$ and introduce the notation 
\begin{eqnarray*}
||\va{f}||^2=\int_{\tau}^{}|\va{f}|^2 d\tau
\end{eqnarray*}
for what is known as the $L_2$ norm of $f$.
Then, after a few integrations by parts that make use of the homogeneous
boundary conditions on $\va{v}$, we have
\begin{eqnarray}
\label{eqdecomp}
\frac{d}{dt} \left(\frac{||\va{v}||^2}{2}\right) = - \intau{\left[\frac{1-\eta}{Re} |\va{\nabla} \va{v}|^2+\va{v}.(\va{\nabla}\va{V}_{lam}).\va{v}\right]} \equiv - \mathcal{H}\{\va{v}\} .
\end{eqnarray}
If the quadratic form $\mathcal{H}$ is strictly positive, i.e., if there is a number $\mu > 0$ such that $\mathcal{H}\{\va{v}\} \ge \mu ||\va{v}||^2$ for any divergence-free field satisfying
the homogeneous boundary conditions, then the kinetic energy of the perturbation $\va{v}$ decreases monotonically in time at least exponentially. 
A flow for which $\mathcal{H}$ is a positive quadratic form is called {\it energy stable}
 --- it possesses a strong form of absolute asymptotic nonlinear stability.

Now we define the function $\mu(Re,\tan \Theta)$ by
\begin{eqnarray}
\label{eqinf} \mu(Re,\tan \Theta)=\inf
\frac{\mathcal{H}\{\va{v}\}}{||\va{v}||^2},
\end{eqnarray}
the infimum (or greatest lower bound) 
being taken over divergence-free vector fields satisfying the homogeneous boundary conditions.   Energy stability is achieved in the region where $\mu(Re,\tan \Theta)>0$ in the $Re- \tan\Theta$ plane. 
The level set $\mu(Re,\tan \Theta)=0$ is the boundary of the parameter region where the flow is marginally energy stable.

\subsection{Euler-Lagrange equations}
\label{Euler-L} 
Only the terms of equation (\ref{NSdecomp}) which are linear in ${\bf v}$ contribute to ${\mathcal{H}}$ so that it is clear that
we can write  $\mathcal{H}\{\va{v}\}=\intau{\va{v} \cdot L\va{v}}$ where $L$ is a symmetric linear operator diagonalizable in an orthonormal basis with real eigenvalues. 
We perform the variational procedure subject to the constraint of incompressibility of the flow by adding the integral $\intau{p\va{\nabla}.\va{v}}$ to ${\mathcal{H}}$. 
A second Lagrange multiplier $\lambda$ is used to impose the constraint $|| \va{v} ||^2=constant$ when seeking the infimum in equation (\ref{eqinf}).  
We may then seek
a stationary condition on the combined functional.    The Euler-Lagrange equations are then
\begin{eqnarray}
\label{ELvect} \lambda \va{v}= - \frac{1-\eta}{Re} \Delta \va{v} +\frac{1}{2}
[(\va{\nabla} \va{V}_{lam}).\va{v}+\va{v}.(\va{\nabla}
\va{V}_{lam})]+\va{\nabla}p
\label{ELeqs}
\end{eqnarray}
together with the incompressibility constraint $\va{\nabla} \cdot \va{v}=0$.
As usual, the Lagrange multiplier, $p$, in the Euler-Lagrange equations plays the role of pressure. The Euler-Lagrange equations for the variational problem are equivalent to the eigenvalue problem for the linear operator $L$ and the infimum in equation (\ref{eqinf}) is reached when ${\va{v}}$ is an eigenvector of $L$ for its lowest eigenvalue. This lowest eigenvalue is the lowest acceptable value of $\lambda$ in equation (\ref{ELvect}).
We stress the fact that although these equations are linear they give sufficient conditions for the laminar flow to be stable against perturbations of arbitrarily large amplitude. 
We solve this problem numerically in \ref{secnumc} but first we consider some analytical aspects. 

In the classical Taylor-Couette flow without suction one can show that for some parameter values, the Taylor-vortices --- that is a $z$-dependent and $\theta$-independent mode --- are linearly stable but may exhibit transient growth (see for instance Joseph, 1976). In the present energy stability analysis of the Taylor-Couette flow with suction, we thus need to retain both $\theta$ and $z$ dependent deviation fields $\va{v}$ to compute the energy stability limit correctly. 

Since the problem is linear and invariant to translations in $\theta$ and $z$, it is useful to decompose $\va{v}$ into Fourier modes
\begin{eqnarray}
\va{v}=(u(r),v(r),w(r)) e^{i m \theta} e^{i k z}.
\end{eqnarray}
The modes decouple so we may study the stability of each one separately.
Insertion of this into the Euler-Lagrange equations yields

\begin{eqnarray}
 0 &=& \left(-\lambda + A(r) + \frac{\tan \Theta}{r^2} \right) u + Z(r) v + p_r - \frac{(1-\eta)}{Re} \frac{1}{r}
 (r u_r)_r \label{ELr}\\
 0 &=& Z^*(r) u + \left(-\lambda + A(r) - \frac{\tan \Theta}{r^2} \right) v + \frac{i m}{r} p -  \frac{(1-\eta)}{Re} \frac{1}{r}(r v_r)_r \label{ELt}\\
 0 &=& \left(-\lambda+ A(r) - \frac{(1-\eta)}{Re} \frac{1}{r^2} \right) w + i k p - \frac{(1-\eta)}{Re} \frac{1}{r}(r w_r)_r \label{ELz}
\end{eqnarray}
where the functions $A$ and $Z$ are
\begin{eqnarray}
A(r) & = & \frac{1-\eta}{Re} \left( k^2 + \frac{m^2+1}{r^2} \right) \\
Z(r) & = & \frac{1}{2}\left(r \left(\frac{V(r)}{r} \right)_r\right)+2 i m \frac{(1-\eta)}{Re} \frac{1}{r^2},
\end{eqnarray}
and $Z^*$ is the complex conjugate of $Z$. The final equation of the system is the incompressibility constraint,
\begin{eqnarray}
\label{mass} \frac{1}{r} (r u)_r + i m \frac{v}{r} + i k w = 0.
\end{eqnarray}
To determine the energy stability of the flow, we need to solve this system of equations together with the homogeneous boundary conditions on $\va{v}$ to find the lowest eigenvalue, $\lambda$. 
 The flow is energy stable if $\lambda>0$ as can be seen by inserting the modal expression into (\ref{eqinf}).

\section{Bounds on the energy stability limit}
\label{secbounds}
In this section we find some bounds on the location of the curve $\mu(Re,\tan \Theta)=0$ in the $Re-\tan\, \Theta$ plane.
To do this, we need not solve the variational problem explicitly, so we instead take two other approaches: 

\begin{itemize}
\item First we specify a particular choice of initial perturbation and we study its energy stability.
If the energy of any disturbance can grow, then the flow is not energy stable.
By identifying such test perturbations we can bound the location of the marginal curve of energy stability from one side. 

\item Then we derive an analytical lower bound on the quadratic form $\mathcal{H}$. 
As long as this lower bound remains positive, $\mathcal{H}$ is positive and the flow is energy stable.
This (rigorous) sufficient condition for the flow to be absolutely stable bounds the location of the marginal curve of energy stability from the other side.
\end{itemize}


\subsection{Stability of the mode $m=k=0$}
\label{m0k0}

Consider variations that are axisymmetric and translationally invariant in the $z$-direction corresponding to the mode $m=k=0$.   
In this case, the mass conservation equation together with the homogeneous boundary conditions on $\va{u}$ yield $u(r)=0$.
To find the marginal energy stability limit we impose $\lambda=0$ in the Euler-Lagrange equations.
After multiplication by $Re/(1-\eta)$ the azimuthal component of the equations simplifies to
\begin{eqnarray}
\label{eqk0m0} v_{rr} + \frac{v_r}{r} + (\alpha - 1)\frac{v}{r^2} = 0
\end{eqnarray}
This equation admits power law solutions with $v(r) \propto r^{\pm (1-\alpha)^{1/2}}$.
We are interested in the case $\alpha>1$ corresponding to the most unstable situation. We thus obtain the general solution.
\begin{eqnarray}
v(r) & = & A \cos (\sqrt{\alpha-1} \ln r) + B \sin(\sqrt{\alpha-1} \ln r).
\end{eqnarray}
The boundary condition $v(1)=0$ implies $A=0$ and $v(\eta) = 0$ then imposes either $B=0$ or, more interestingly, $\sqrt{\alpha-1} \ln(\eta) = q \pi$, where $q$ is an integer.
As $\alpha$ increases, the first mode to leave the domain of  energy stability has $q=1$ for which the corresponding critical value of 
$\alpha$ is $\alpha_c=1+\frac{\pi^2}{(\ln\, \eta)^2}$.
On returning to the $Re - \tan\Theta$ plane, we find an upper bound on the critical value of the 
Reynolds number at a given value of $\tan \Theta$, namely
\begin{eqnarray}
Re_c(\tan\, \Theta) &\le& Re_1(\tan\, \Theta) = \left( 1+\frac{\pi^2}{(\ln(\eta))^2} \right) \frac{1-\eta}{\tan \Theta}.
\end{eqnarray}
The curve $Re_1(\tan(\Theta))$ has been drawn on Figure \ref{ESlimitfig} for several values of the geometrical factor $\eta$. 
This upper bound shows that whatever the injection angle is, energy stability will be lost when the Reynolds number becomes high enough.
 
\begin{figure}[h]
\begin{center}
\includegraphics[width=130 mm]{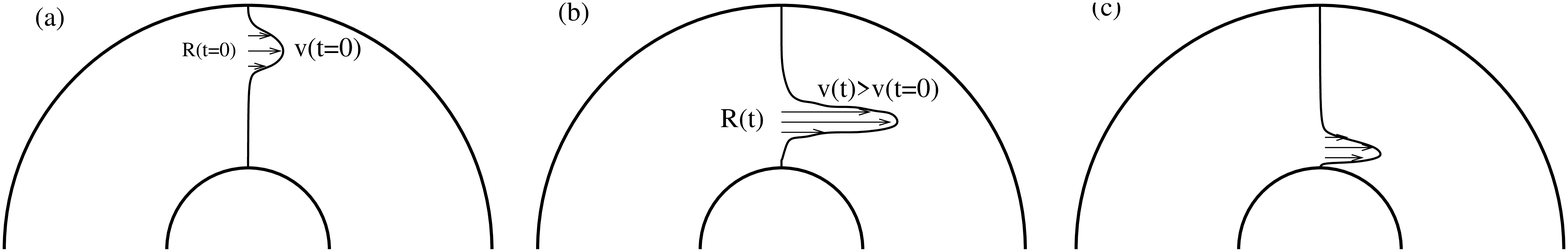}
\caption{Expected behavior of an energy unstable mode for $m=k=0$: a perturbation at $t=0$ (a) is advected by the suction and speeds up (b). When it reaches the boundary layer the fluid loses its angular momentum and the perturbation is swept away (c).}
\label{unst}
\end{center}
\end{figure}

Figure \ref{unst} illustrates how the $m=0$, $k=0$ mode may grow---at least transiently--- when suction spoils the conventional interchange argument for large values of $\alpha$.
A small perturbation $v_0=v(t=0)$ in the azimuthal velocity field at an initial radius $R_0=R(t=0)$ will be advected towards the center by the suction.
When $\alpha$ is very large, the flow is nearly inviscid and the fluid nearly conserves its angular momentum during the inward motion. 
Hence the velocity perturbation becomes 
\begin{eqnarray}
v(R(t),t)=v_0 \frac{R_0}{R(t)}, \mbox{ with } R(t) \leq R_0,
\end{eqnarray}
which means that the kinetic energy of this perturbation increases and the mode is not energy stable.
However, when the perturbation is advected all the way to the boundary layer near the inner cylinder, viscous effects dissipate the angular momentum of the perturbation and the perturbation is removed by the suction.
\bigskip
\bigskip

\subsection{Lower bound on the energy stability limit}
\label{lowbound} 
Write the quadratic form $\mathcal{H}\{\va{v}\}$ as
\begin{eqnarray}
\mathcal{H}\{\va{v}\} = \intau{\frac{1-\eta}{Re} |\va{\nabla} \va{v}|^2} + \mathcal{F}\{\va{v}\}
\end{eqnarray}
with the indefinite term
\begin{eqnarray}
\mathcal{F}\{\va{v}\}=\intau{ \left( r \left( \frac{V(r)}{r} \right)_r \right) uv - \frac{\tan \Theta}{r^2}(v^2-u^2)}.
\end{eqnarray}
To find a lower bound on $\mathcal{H}$ we observe that $\va{v}$ cannot be large within the gap while satisfying the homogeneous boundary conditions without also having large gradients. 
Thus we seek a lower bound on $\mathcal{F}$ (which may be negative) in terms of the non-negative norm $||\va{\nabla} \va{v}||^2$. 

First of all, for a lower estimate we can drop the positive term $\frac{\tan \Theta}{r^2} u^2$ in $\mathcal{F}$. 
Then using the inequality $|uv| \leq \frac{1}{2}[\frac{1}{c} u^2 + c v^2]$, valid for any $c>0$, we have
\begin{eqnarray}
\mathcal{F}\{\va{v}\} \geq -\intau{u^2
\left(\frac{|r(V/r)_r|}{2 c} \right) + v^2 \left( \frac{\tan \Theta}{r^2}
+ c \frac{|r(V/r)_r|}{2}\right)}.
\end{eqnarray}
Now choose $c=c(r)$ so that the coefficients of $u^2$ and $v^2$ are equal:
\begin{eqnarray}
c(r)=\frac{\chi}{2} \left(\sqrt{1+\frac{4}{\chi^2}}-1\right)
\end{eqnarray}
where $\chi(r) =  \frac{2 \tan \Theta}{r^3|(V/r)_r|}$.
This choice implies
\begin{eqnarray}
\label{low} 
\mathcal{F}\{\va{v}\} \geq - \intau{\frac{\tan \Theta}{2r^2} \left(\sqrt{1+\frac{4}{\chi^2}}+1 \right) (u^2+v^2) }. \quad
\end{eqnarray}
We now use the fundamental theorem of calculus and the Schwartz inequality to find
\begin{eqnarray}
|v(r)| = \left| \int_{R_1}^{r} \frac{1}{\sqrt{\tilde{r}}}
\sqrt{\tilde{r}}  v_r(\tilde{r}) d\tilde{r} \right| \nonumber \leq \sqrt{\ln \left( \frac{r}{R_1} \right)}
\sqrt{\int_{R_1}^{r} \left|v_r(\tilde{r})\right|^2 \tilde{r}
d\tilde{r}}. \quad
\end{eqnarray}
The same estimate can be applied to $u$ and yields
\begin{eqnarray}
u^2+v^2 \leq \ln \left( \frac{r}{R_1} \right) \int_{R_1}^{r} \left(
\left|u_r(\tilde{r})\right|^2 + \left|v_r(\tilde{r})\right|^2\right)
\tilde{r} d\tilde{r}. \quad
\end{eqnarray}
Now we use this estimate in (\ref{low}) to find
\begin{eqnarray}
\mathcal{F}\{\va{v}\} &\geq& - ||\va{\nabla} \va{v} ||^2 \int_{R_1}^{R_2} \frac{\varphi}{2r} \ln
\left( \frac{\tilde{r}}{R_1} \right) \left(
\sqrt{1+\frac{4}{\chi^2}}+1 \right) d\tilde{r}. \quad 
\end{eqnarray}
Thus $\mathcal{H}\{\va{v}\}=\frac{1-\eta}{Re} ||\va{\nabla} \va{v}||^2+\mathcal{F}\{\va{v}\} \geq 0$ if
\begin{eqnarray}
\frac{1-\eta}{Re} -  \int_{\eta}^{1} \frac{\tan \Theta}{2\tilde{r}} \ln \left(
\frac{\tilde{r}}{\eta} \right) \left( \sqrt{1+\frac{4}{\chi^2}}+1
\right) d\tilde{r} \geq 0. \quad
\end{eqnarray}

If we compute $\chi(r)$ for the laminar azimuthal velocity profile found in section \ref{lamsol}, the lower bound on the energy stability limit becomes the implicit relation
\begin{eqnarray}
\frac{\alpha}{2} \int_{\eta}^{1} \frac{\ln \left( \frac{\tilde{r}}{\eta}
\right)}{\tilde{r}} \left( \sqrt{1+\frac{4}{\chi^2}}+1
\right) d\tilde{r} =  1
\end{eqnarray}
with
\begin{eqnarray}
\chi(r)  =  2 \tan{\Theta}
\frac{|\eta^{2+\alpha}-1|}{|\alpha r^{2+\alpha} + 2
\eta^{2+\alpha}|} \ .
\end{eqnarray}
The line $Re_2(\tan\, \Theta)$ corresponding to this bound is shown in Figure \ref{ESlimitfig} for several values of $\eta$.
Below this curve, the flow is absolutely stable, so the actual marginal energy stability boundary lies above it.

\section{Numerical computation of the energy stability limit} 
\label{secnumc}

In a study of plane Couette flow with suction, Doering {\it et al.}  (2000) found that steady laminar flow was absolutely stable if the injection angle was above the critical value $\Theta_c \simeq 3^o$. 
At this value of the injection angle the energy stability boundary in the $Re-\, \tan\, \Theta$ plane goes to infinity vertically. 
In the cylindrical problem, however, the upper bound found in \ref{m0k0} clearly rules out such a behavior.
It is interesting to see how the energy stability boundary evolves from the plane Couette limit $\eta \rightarrow 1$ to a cylindrical geometry with $\eta < 1$.
To answer this question, we solved the eigenvalue problem in (\ref{ELeqs}) numerically to determine the marginal energy stability curve precisely.

\subsection{Simplification of the system of equations}

Although the system of equations (\ref{ELr})-(\ref{mass}) shares some similarity with the system of equations (\ref{lin1})-(\ref{linmass}) studied in section \ref{seclin}, the presence of $z$-dependence makes its simplification less straightforward. 
Taking the divergence of the vectorial form of the Euler-Lagrange equation and using the incompressibility constraint we obtain
\begin{eqnarray}
\label{deltap}  \Delta p = -\frac{1}{2 r} \left( r^2 \left( V/r \right)_r v  \right)_r - \frac{i m}{2 r} \left( r \left( V/r \right)_r \right) u + \frac{\tan \Theta}{r^2} \left[-u_r + \frac{u}{r} + \frac{i m v}{r}\right].
\end{eqnarray}
Next we apply the Laplacian, $\Delta$, to (\ref{ELr}) and use the identity
\begin{eqnarray}
\Delta(p_r)=(\Delta p)_r + \frac{p_r}{r^2} - \frac{2 m^2}{r^3} p
\end{eqnarray}
to eliminate $\Delta(p_r)$ from the result.
The remaining terms involving $p$ are $(\Delta p)_r$, $p_r$ and $p$, which can be expressed in terms of $u$, $v$ and their derivatives using (\ref{deltap}), (\ref{ELr}) and (\ref{ELt}).
This leads to the first of a system of two differential equations for $u$ and $v$, which is
\begin{eqnarray}
& & \lambda \left[ \frac{-2 i m}{r^2} v + u_{rr} + \frac{1}{r} u_r -
(k^2 +\frac{m^2+1}{r^2}) u \right]  \nonumber = \left( -\frac{1-\eta}{Re} \right) u_{rrrr} + \left( -\frac{2 (1-\eta)}{Re~r} \right) u_{rrr}  \\
\nonumber & & + \left( 2 A + \frac{1-\eta}{Re~r^2} \right) u_{rr}+ \left( \frac{2 A}{r} -\frac{1-\eta}{Re~r^3} + 2A_r -
\frac{i m (V/r)_r}{2} \right) u_r \\
\nonumber & & + \left( A_{rr} + \frac{A_r}{r} - A\left(k^2+\frac{m^2+1}{r^2}\right) - \frac{i m (V/r)_{rr}}{2} -\frac{2 i m}{r^2} Z^* \right) u + \left( \frac{4 i m (1-\eta)}{Re~r^2} \right) v_{rr}   \\
\nonumber & & + \left( -\frac{4 i m (1-\eta)}{Re~r^3} \right) v_r + \left(\frac{6 i m (1-\eta)}{Re~r^4} -Z(k^2+\frac{m^2}{r^2}) - \frac{2 i m A}{r^2}\right) v \\
& & + \tan \Theta \left[ - \left( \frac{k^2}{r^2} + \frac{m^2}{r^4} \right) u + \frac{i m}{r^3} v_r - \frac{i m}{r^4} v \right] .
\end{eqnarray}
\newline
To derive the second equation of the system, we express $w$ in terms of $u$ and $v$ using mass conservation (\ref{mass}).
On inserting $w$ into (\ref{ELz}), we find an expression for $p$ that can be inserted into (\ref{ELt}).
This leads to
\begin{eqnarray}
\nonumber & & \lambda \left[ -\frac{i m}{r} u_r - \frac{i m}{r^2} u +
(k^2+\frac{m^2}{r^2}) v \right] =  \left( \frac{i m (1-\eta)}{Re~r}\right) u_{rrr} + \left( \frac{2 i m (1-\eta)}{Re~r^2}\right) u_{rr}\\
\nonumber & & + \left( -
\frac{i m}{r} A \right) u_{r} + \left( k^2 Z^* - \frac{i m}{r^2} A + \frac{2 i m
(1-\eta)}{Re~r^4} \right) u \\
\nonumber & & + \left( -\frac{1-\eta}{Re} \left(k^2+\frac{m^2}{r^2} \right)\right)
v_{rr} + \left(\frac{1-\eta}{Re~r} \left( - k^2+\frac{m^2}{r^2}\right)\right) v_{r} \\
& & + \left( A (k^2+\frac{m^2}{r^2}) - \frac{2 (1-\eta)
m^2}{Re~r^4} + \frac{k^2  \tan \Theta}{r^2}\right) v
\end{eqnarray}
\newline
This system is of fourth order in $u$ and second order in $v$ and there are, appropriately, four boundary conditions on $u$ and two on $v$.

\subsection{From plane Couette to cylindrical Couette}

\begin{figure}[]
\begin{center}
\includegraphics[width=70 mm]{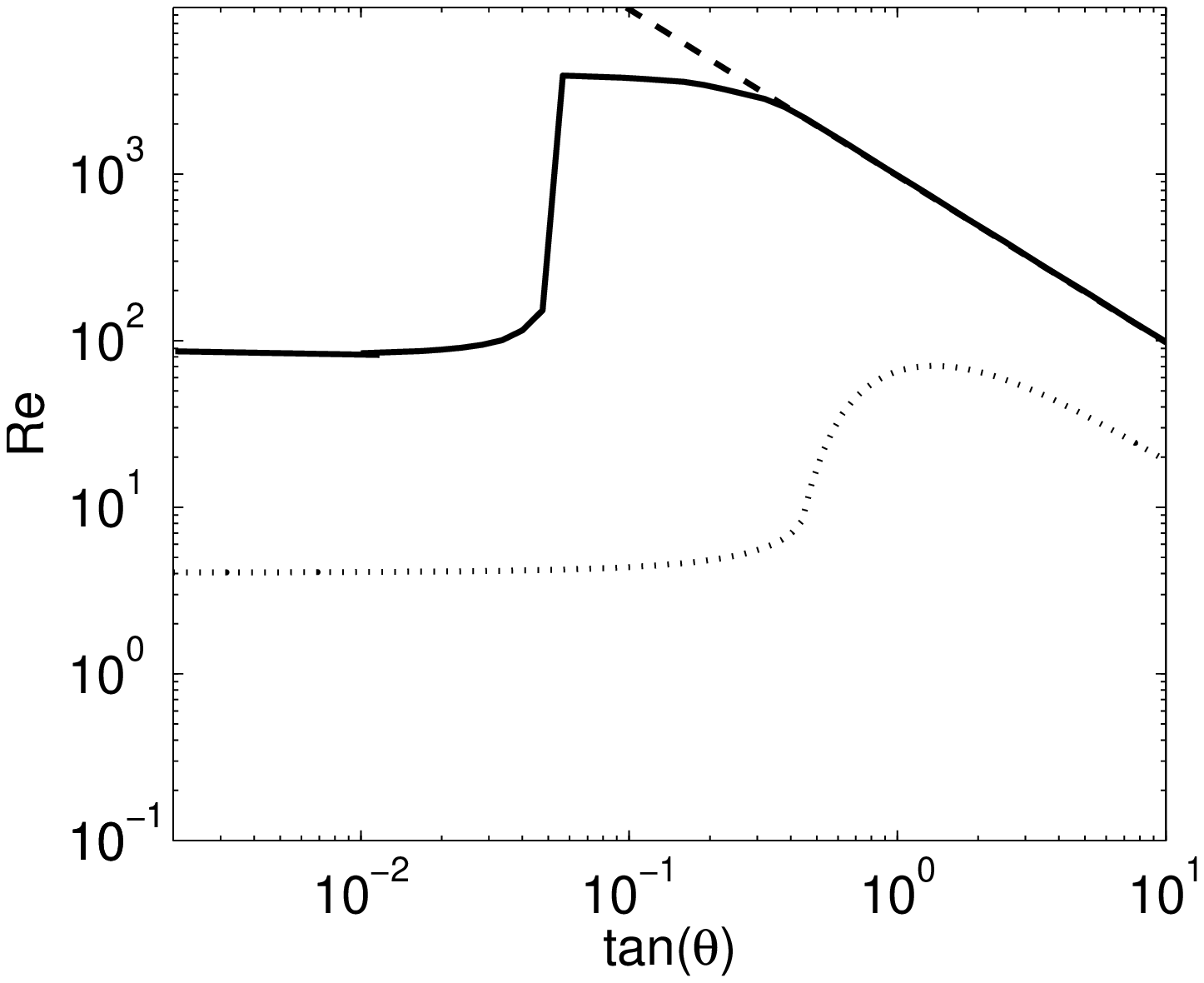}\newline
\includegraphics[width=70 mm]{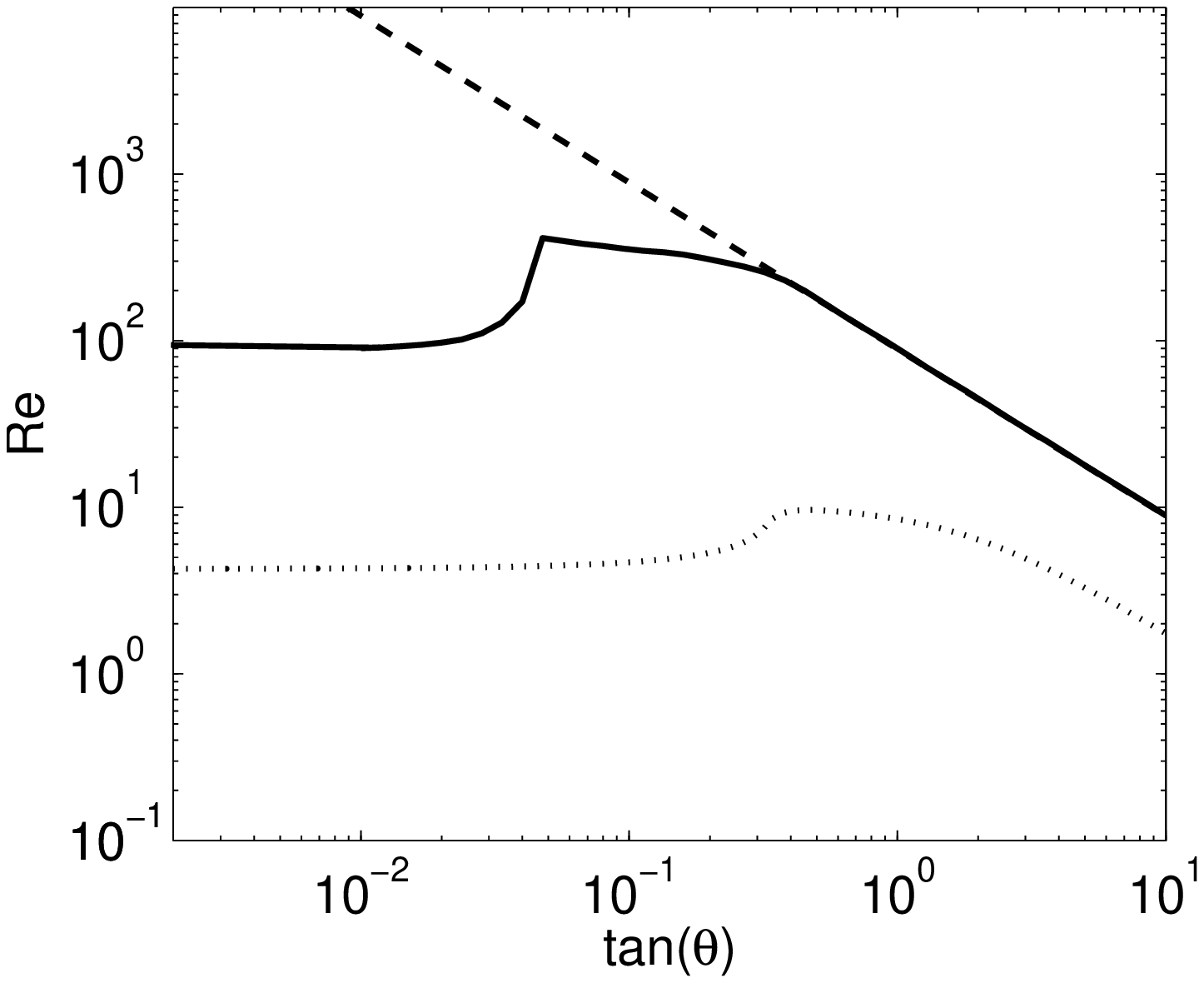}\newline
\includegraphics[width=70 mm]{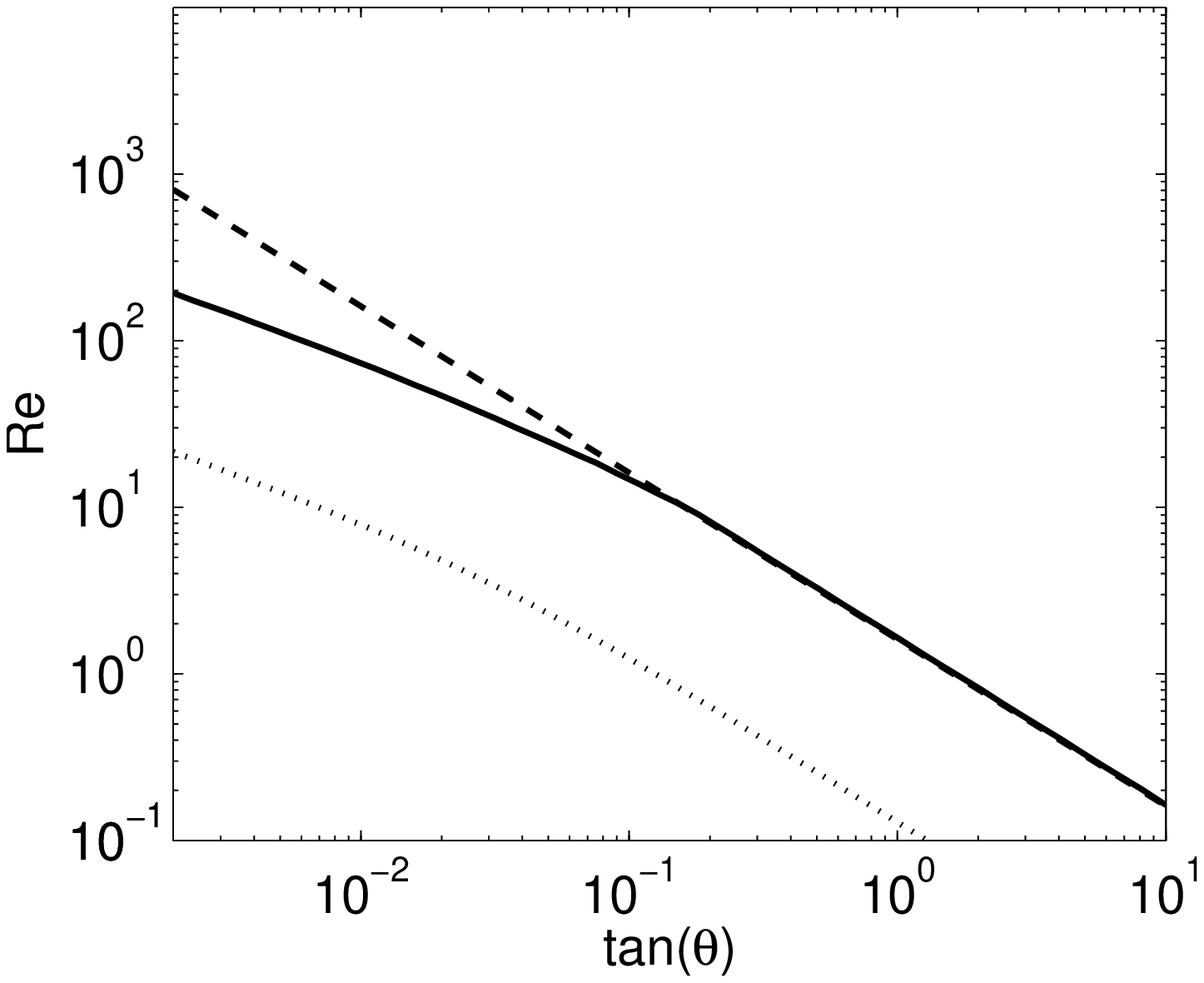}\newline
\caption{Marginal energy stability boundary for different values of $\eta$ (top: $\eta=0.99$, middle: $\eta=0.9$, bottom: $\eta=0.02$).
The solid line is the numerically-computed energy stability boundary, the dashed line is the upper bound $Re_1$ on the location of that curve, and the dotted line is the lower bound $Re_2$.}
\label{ESlimitfig}
\end{center}
\end{figure}

We solved the eigenvalue problem numerically using a finite-difference method.
The resulting marginal energy stability curves are shown for several values of $\eta$ in Figure \ref{ESlimitfig}. 
When $\eta$ is close to one and the entry angle is very small,
the energy stability limit remains constant at the energy stability limit of the plane Couette flow, $Re\simeq82$.
When $\theta \simeq 3^o$ the energy stability limit increases greatly but it cannot go to infinity as in the plane geometry. 
When $0.3 \lesssim \tan\, \Theta$, the most unstable 
mode is $(m=0, k=0)$ and the energy stability limit coincides with the upper bound $Re_1$.

When $\eta$ goes to zero, the energy stability boundary is very different from that of the plane case.  This limit corresponds to the situation where the outer radius goes to infinity while the inner radius is kept constant, and the boundary curve is a monotonically decreasing function of the injection angle.  The critical Reynolds number at $\Theta=0$ is higher in this case.
\section{Conclusion}
\label{conclusion}
It is not unusual to suspect that a given flow may be unstable only to encounter mathematical arguments to the contrary.
A familiar example for fluid dynamicists is provided by kinematic dynamo theory.
In a working kinematic dynamo, a magnetic field grows exponentially from a small magnetic perturbation on a suitable flow.  To many this prospect once seemed to be forbidden by Cowling's  antidynamo theorem.
The way out of the dilemma was to (finally) appreciate that Cowling's theorem posited an axisymmetric flow and to focus on flows that did not respect this constraint.
In the present problem we confront a similar situation: many flows with differential rotation may seem to have enough energy to excite the growth of disturbances to the flow yet Rayleigh's criteria appear to forbid a purely fluid dynamical instability.
In this instance, we have contravened Rayleigh's proscription by simply violating one of the conditions of Rayleigh's demonstrations.
In both of these examples, the value of the relevant (anti-)theorem has been to point the way to making the (seemingly) forbidden event happen by vitiating one of the theorem's
premises. 

As we mentioned at the outset, there are several interesting problems where differential rotation plays an important role.
We have not hidden our particular interest in the situation of accretion disks where matter rotates differentially around a central gravitating body.
These disks are often quite luminous and the source of their emissions has been thought to be in the dissipation of the turbulence that they had long been assumed to sustain.
But these objects typically have Keplerian rotation laws where their circular velocities vary as $1/\sqrt{r}$, where $r$ is the distance from the central object.
Hence there was no generally accepted purely fluid dynamical process to produce this turbulence, though the matter was steeped in controversy.
Moreover, turbulence in the disk, if it occurred, was supposed to transport angular momentum out of the disk.
This would allow the purely circular motion to begin to turn inward so that the gas in the disk could fall onto the central object and be accreted.
The resulting spiral motion, or swirl, is then a central feature of the accretion disks yet its dynamical role has been largely glossed over in the discussions of disk instabilities.   

For us, the intriguing feature of the instability of swirling (a.k.a. spiraling) flows in disks is that their instabilities may promote accretion that may in turn enable the inward flow 
that was the source of instability. This is positive feedback {\bf par excellence} and we may reasonably anticipate that the instability treated here is likely to cause subcritical bifurcation in disks, though we have not yet shown that.
Nevertheless, this physical argument leads us to suppose that rather than thinking of a competition among instability mechanisms, we may imagine a cooperation amongst them.
Thus in the linear study of the magneto-rotational instability Kersale {\it et al.} (2004) find that the resulting fluid stresses drive accretion for suitable boundary conditions.
This self-induced accretion means that a swirl is produced that is likely to contribute to the instability we consider here.  This interaction of instabilities in the fully nonlinear regime bids fair to give rise to many of the processes that have been sought to explain the observed wealth of behavior in swirling flows.

However, the work reported here does not treat disks explicitly.
We have chosen instead to consider a fluid flow field that is well known to be stable and to show that by converting it from purely circular to swirling motion by the agency of accretion we could render it linearly unstable.
This required only a simple change in the boundary conditions to vitiate the applicability of Rayleigh's inflection point theorem for circular motion.
It remains to learn whether this destabilization will apply to Keplerian flows, though the mechanism seems generic and we do think the instability is inherent to physically
realistic, spiraling flows.
As to real disks, there we are in the domain of shallow gas theory and that will lead to a more difficult investigation, but in the beginning it will be a linear one.
In the  meantime, it could well be that other instabilities will be uncovered as other variations on disk flows are examined.  These will also join the magnetorotational instability that now leads the pack and whose consequences are currently being vigorously explored by many investigators.
 \break \bigskip

\noindent
{\it Acknowledgment.} This work was initiated, and much of it carried out, during the 2007 Geophysical Fluid Dynamics Program at Woods Hole Oceanographic Institution.
We are grateful to the US National Science Foundation and the US Office of Naval Research for their ongoing support of this program.
This work was also supported in part by NSF Awards PHY-0555324 and PHY-0855335.

\appendix

\section{Bounds on the energy dissipation}
\label{secbounde}
The first physically relevant rigorous bound for turbulent flow was given by Howard (1963) 
using an approach to thermal convection proposed by Malkus (1960).
Subsequently, related techniques have been developed such as the background method (Doering and Constantin, 1994, 1996). 
Although the background method can be applied to many problems in many geometries, there has been much emphasis on shear layers.  
For example Constantin (1994) used the method to compute a bound on the energy dissipation for the Taylor-Couette flow without suction and Doering, Spiegel \& Worthing (2000) found a bound on the energy dissipation in a plane shear layer with suction. 
However, at high Reynolds numbers and strong suction the Taylor-Couette problem appears to be significantly more challenging.
We show here that the usual formulation of the background method is inadequate in this case.

\subsection{The background method}

The background method begins with a decomposition of the velocity field into a steady velocity profile and a time-dependant fluctuation field $\va{u(\va{r},t)}=\va{V(\va{r})}+\va{v(\va{r},t)}$. 
The difference from energy stability analysis is that the background profile remains to be chosen; it is not necessarily the basic steady solution.   
Moreover it needs not be a mean field as in the Malkus-Howard approach.
We insert this decomposition into equation (\ref{NS}), take the dot product with $\va{v}$ and integrate over the volume $\tau$.
We perform some integrations by parts and use the equality
\begin{eqnarray}
|\va{\nabla} \va{u}|^2=|\va{\nabla} \va{V}|^2+|\va{\nabla}
\va{v}|^2+2(\va{\nabla} \va{v})\cdot(\va{\nabla} \va{V})
\end{eqnarray}
to deduce
\begin{eqnarray}
\label{eqbtemp} 
\frac{d}{dt}\left(\frac{||\va{v}||^2}{2} \right) +
\frac{1-\eta}{2 Re} ||\va{\nabla} \va{u}||^2  =  \frac{1-\eta}{2 Re}
||\va{\nabla} \va{V}||^2 - I 
\end{eqnarray}
with
\begin{eqnarray}
 I = \intau{\left[\frac{1-\eta}{2 Re}|\va{\nabla}
\va{v}|^2+\va{v}.(\va{\nabla}
\va{V}).\va{v}+\va{v}.(\va{V}.\va{\nabla}).\va{V}\right]}. \ \ \ 
\end{eqnarray}
The last term in $I$ is linear in $\va{v}$ which is not desirable in this procedure so, to alleviate this problem, we introduce another decomposition,
\begin{eqnarray}
\va{v}=\va{W}(\va{r})+\va{w}(\va{r},t)
\end{eqnarray}
where $\va{W}$ and $\va{w}$ are both divergence-free and satsify the homogeneous boundary conditions. 
($\va{W}$ is steady whereas $\va{w}$ may be time dependent.) 
This decomposition is inserted in $I$ to yield

\begin{eqnarray}
& & I = \mathcal{I}\{\va{w}\}+\frac{1-\eta}{2 Re} ||\va{\nabla} \va{W}||^2 + \intau{\left[\va{W}\cdot(\va{\nabla} \va{V})\cdot\va{W} +
\va{W}\cdot(\va{V}\cdot\va{\nabla})\cdot\va{V}\right]} + \mathcal{L}\{\va{w}\}
\end{eqnarray}
where
\begin{eqnarray}
\mathcal{I}\{\va{w}\} & = & \intau{\left[\frac{1-\eta}{2 Re}
|\va{\nabla}\va{w}|^2 +
\va{w}\cdot(\va{\nabla} \va{V})\cdot\va{w}\right]} \nonumber \\
\mathcal{L}\{\va{w}\} & = & \intau{ \va{w}\cdot [ -\frac{1-\eta}{Re} \Delta \va{W}
+ (\va{\nabla} \va{V})\cdot\va{W} + \ (\va{W}\cdot\va{\nabla})\va{V} +
(\va{V}\cdot\va{\nabla})\va{V} ] }.
\end{eqnarray}
If the quadratic form $\mathcal{I}$ is positive, a field $\va{W}$ exists that may be employed to remove the linear term in $\va{w}$ ($\mathcal{I}>0$ is an invertability condition for the linear operator acting on $\va{W}$ inside the square brackets in $\mathcal{L}$).
Then $\mathcal{L}\{\va{w}\}=0$ for any vector field $\va{w}$ that is divergence-free and satisfies the homogeneous boundary conditions
(this does not mean that the bracket inside $\mathcal{L}$ has to be zero, just that it is a gradient). 
When such a vector field $\va{W}$ is found, then $\mathcal{L}\{\va{W}\}=0$ as well, and
\begin{eqnarray}
\mathcal{I}\{\va{W}\}=-\frac{1}{2}
\intau{\va{W}\cdot(\va{V}\cdot\va{\nabla})\va{V}}.
\end{eqnarray}
The quantity $I$ is then
\begin{eqnarray}
I = \mathcal{I} \{\va{w}\} + \frac{1}{2}
\intau{\va{W}\cdot((\va{V}\cdot\va{\nabla})\va{V})},
\end{eqnarray}
which when inserted in equation (\ref{eqbtemp}) yields

\begin{eqnarray}
d_t(||\va{v}||^2) &+& \frac{1-\eta}{Re} ||\va{\nabla} \va{u}||^2 = \frac{1-\eta}{Re} ||\va{\nabla} \va{V}||^2 - \intau{\va{W}\cdot((\va{V}\cdot\va{\nabla})\va{V})} - 2 \mathcal{I}\{\va{w}\}. \quad \quad
\label{BG1}
\end{eqnarray}
$\mathcal{I}\{\va{w}\}$ is the same quadratic form as for the energy stability analysis of $\va{V}$ as if it were a steady solution with $Re$ replaced by $2 Re$. 
If the background velocity field $\va{V}$  may be chosen so that $\mathcal{I}$ is a positive quadratic form, then the corresponding $\va{W}$ exists and $\mathcal{I}\{\va{w}\}$ can be dropped in (\ref{BG1}) so that
\begin{eqnarray}
\frac{1-\eta}{Re} \overline{||\va{\nabla} \va{u}||^2} \leq  \frac{1-\eta}{Re} ||\va{\nabla}
\va{V}||^2 - \intau{\va{W}\cdot(\va{V}\cdot\va{\nabla})\va{V}} \quad
\label{BGbound}
\end{eqnarray}
where the overline represents a long time average.

In most applications, background profiles are chosen to depend on the same coordinates as the simplest steady laminar solutions that respect the symmetries of the problem.
By choosing the background as a function of the depth only, 
we may concentrate the background shear into thin layers near the no-slip boundaries. Background flows can then be found that ensure that $\mathcal{I}$ is a positive quadratic form, at the cost of increasing the bound on the right hand side of (\ref{BGbound}).
Choosing an azimuthal background velocity profile that depends only on $r$ as for the laminar solution is a successful strategy for the classical Taylor Couette problem (Constantin, 1994).

In the Taylor-Couette flow with suction, there is a limit on the energy stability domain of any background flow that cannot be extended:
the energy stability boundary that emerges from the study of the $(m=0,k=0)$ mode, does not involve the azimuthal velocity profile.
This means that for any injection angle $\Theta$, and for any Reynolds number greater than $2 Re_1(\Theta,\eta)$, the quadratic form $\mathcal{I}$ will not be a positive quadratic form no matter what function of $r$ is chosen to be the background azimuthal velocity profile.
Therefore we cannot produce a bound on the energy dissipation with a ``simple'' background azimuthal velocity depending only on $r$.
(This sort of mathematical obstruction was also recently discovered in an abstract setting by Kozono and Yanagisawa (2009).)  There is no doubt a message in this limitation on the
method but we shall not speculate on this here.  

In studies on convection or shear layers, optimized background profiles resemble mean flow profiles at high Rayleigh and Reynolds numbers (Doering and Constantin, 1994 and 1996). 
This reinforces the common belief that the symmetries of the problem which are lost through symmetry-breaking bifurcations are recovered in a statistical sense for developed turbulence at even higher Reynolds numbers. 
The fact that we cannot choose a background velocity profile for the Taylor-Couette flow with suction that respects the symmetries of the problem raises the question of whether or not these symmetries may be recovered in any sense at high Reynolds number.

\end{document}